\documentclass[twocolumn,showpacs,nofootinbib,aps,superscriptaddress,prd]{revtex4}
\usepackage{graphicx,dcolumn,bm,amsfonts,amsmath,color,xcolor}
\usepackage[colorlinks=true,citecolor=blue, linkcolor=blue,urlcolor=blue]{hyperref}

\allowdisplaybreaks
\begin{document}

\title{An improved light-cone harmonic oscillator model for the pionic leading-twist distribution amplitude}
\author{Tao Zhong}
\email{zhongtao1219@sina.com}
\address{Department of Physics, Guizhou Minzu University, Guiyang 550025, P.R. China}
\author{Zhi-Hao Zhu}
\address{College of Physics and Materials Science, Henan Normal University, Xinxiang 453007, P.R. China}
\author{Hai-Bing Fu\footnote{Corresponding author}}
\email{fuhb@cqu.edu.cn}
\address{Department of Physics, Guizhou Minzu University, Guiyang 550025, P.R. China}
\address{Department of Physics, Chongqing University, Chongqing 401331, P.R. China}
\author{Xing-Gang Wu}
\email{wuxg@cqu.edu.cn}
\address{Department of Physics, Chongqing University, Chongqing 401331, P.R. China}
\address{Chongqing Key Laboratory for Strongly Coupled Physics, Chongqing 401331, P.R. China}
\author{Tao Huang}
\email{huangtao@ihep.ac.cn}
\address{Institute of High Energy Physics and Theoretical Physics Center for Science Facilities, Chinese Academy of Sciences, Beijing 100049, P.R. China}

\date{\today}
\begin{abstract}
In this paper, we study the pion leading-twist distribution amplitude $\phi_{2;\pi}(x,\mu)$ by improving the traditional light-cone harmonic oscillator model within the reconstruction of the function $\varphi_{2;\pi}(x)$. In order to constraining the model parameters, we calculate its moments $\langle\xi^n\rangle_{2;\pi}|_\mu$ in the framework of QCD background field theory sum rule (BFTSR) up to $10^{\rm th}$ order. Considering the fact that the sum rule of the $0^{\rm th}$ moment $\langle\xi^0\rangle_{2;\pi}|_\mu$ cannot be normalized, we suggest a more reasonable sum rule formula for $\langle\xi^n\rangle_{2;\pi}|_\mu$. Then, we obtain the values of $\langle\xi^n\rangle_{2;\pi}|_{\mu_0}$ with $n=(2,4,6,8,10)$ at the initial scale $\mu_0 = 1~{\rm GeV}$. The first two moments are: $\langle \xi^2\rangle_{2;\pi}|_{\mu_0}  = 0.271 \pm 0.013$, $\langle\xi^4\rangle_{2;\pi}|_{\mu_0} = 0.138 \pm 0.010$; and the corresponding Gegenbauer moments are $a^{2;\pi}_2(\mu_0) = 0.206 \pm 0.038$, $a^{2;\pi}_4(\mu_0) = 0.047 \pm 0.011$, respectively. After fitting the moments $\langle\xi^n\rangle_{2;\pi}|_{\mu}$, we obtained the appropriate model parameters by using the least square method. The resultant behavior for twist-2 pion DA is more closely to the AdS/QCD and lattice result, but is narrower than that by Dyson-Schwinger equation. Furthermore, we calculate the pion-photon transition form factors (TFF) and $B\to\pi$ TFF within light-cone sum rule approach, which are conform with experimental and theoretical results.
\end{abstract}

\pacs{12.38.-t, 12.38.Bx, 14.40.Aq}
\maketitle

\section{introduction}
Light meson light-cone distribution amplitudes (DAs) are universal nonperturbative objects, which describe the momentum fraction distributions of partons in a meson for a particular Fock state. Those DAs enter exclusive processes based on the factorization theorems in the perturbative QCD theory (pQCD), and therefore they are key parameters in the QCD predictions for corresponding processes. In the standard treatment of exclusive processes in QCD proposed by Brodsky and Lepage~\cite{BL}, cross sections are arranged according to different twist structures of meson DAs. In which the leading-twist DA contribution usually dominates due to the contributions from the higher twists are high power suppressed at short distance. Thereafter, the study of pionic leading-twist DA, which describes the momentum distribution of the valence quarks in pion, has attracted much attention in the literature.

So far, a large number of studies on the pionic leading-twist DA rely on its Gegenbauer expansion series~\cite{Lepage:1979zb, Efremov:1979qk}, the nonperturbative expansion coefficients, denoted $a_n^{2;\pi}(\mu)$, are called Gegenbauer moments which encode the long-distance dynamics at low energy scale ($\sim 1 {\rm GeV}$), and only the even Gegenbauer moments are nonzero due to the isospin symmetry. In many applications of pion leading-twist DA involving a high normalization scale, the higher Gegenbauer moment contributions are suppressed due to the anomalous dimension of $a_n^{2;\pi}(\mu)$ grows with $n$, and only the lowest Gegenbauer moments are retained. Therefore, people usually adopt the truncated form involving only the first few terms in the Gegenbauer expansion series to be an approximate form of $\phi_{2;\pi}(x,\mu)$. Those Gegenbauer moments can be calculated directly via some non-perturbative methods such as QCD sum rules~\cite{Chernyak:1981zz, Chernyak:1983ej, Chernyak:1984bm, Huang:1984nc, Xiang:1984hw, Huang:1986wm} or lattice gauge theory~\cite{Gottlieb:1985bn, Gottlieb:1986ie, Martinelli:1988xs, Daniel:1990ah} and so on. Using QCD sum rules to calculate $a_n^{2;\pi}(\mu)$ is realized by calculating $\langle\xi^n\rangle_{2;\pi}|_{\mu}$. Recently, we realized that these calculations need to be improved. By QCD sum rules method, the analytic formula is for $\langle\xi^n\rangle_{2;\pi}|_{\mu} \times \langle\xi^0\rangle_{2;\pi}|_{\mu}$, but which is usual seen as the sum rules of $\langle\xi^n\rangle_{2;\pi}|_{\mu}$ due to normalization of $\phi_{2;\pi}(x,\mu)$. In fact, due to the incompleteness of our sum rules calculation, the deviation of $\langle\xi^0\rangle_{2;\pi}|_{\mu}$ from normalization must be considered. This motivates us to recalculate the moments of pionic leading-twist DA with QCD sum rules.

On the other hand, the truncated form mentioned above does not seem to be enough to describe the behavior of DA under the low energy scale. A natural idea is to consider the contributions of higher-order Gegenbauer polynomials, which requires the calculation of higher-order Gegenbauer moments. But there is a very serious difficulty in doing so, that is, it is difficult to get reliable higher $a_n^{2;\pi}(\mu)$. Through the mathematical relationship between $a_n^{2;\pi}(\mu)$ and $\langle\xi^n\rangle_{2;\pi}|_{\mu}$, we can find that with the increase of order $n$, the reliability of $a_n^{2;\pi}(\mu)$ decreases sharply, which makes our calculation of higher $a_n^{2;\pi}(\mu)$ meaningless. So people try to study the behavior of $\phi_{2;\pi}(x,\mu)$ through other ways. In Ref.~\cite{Chang:2013pq}, in the framework of Dyson-Schwinger equations the authors obtain the pionic leading-twist DA (DS model) that are concave and significantly broader than the asymptotic DA. Making use of the approximate bound state solution of a hadron in terms of the quark model as the starting point, Brodsky-Huang-Lepage (BHL) suggest the light-cone harmonic oscillator model (LCHO model) which is obtained by connecting the equal-time wavefunction (WF) in the rest frame and the WF in the infinite momentum frame~\cite{BHL}. Meanwhile, the holographic Schr\"odinger equation for meson maps onto the fifth dimension of anti-de Sitter with QCD potential (AdS/QCD)~\cite{Ahmady:2017jgq}.

In this paper, we will study the pionic leading-twist DA $\phi_{2;\pi}(x,\mu)$ based on the improved LCHO model. The determination of model parameters depends on the moments $\langle\xi^n\rangle_{2;\pi}|_{\mu}$ rather than the Gegenbauer moments $a_n^{2;\pi}(\mu)$, and we will adopt a new method, that is, the least square method fitting, to determine the model parameters directly. Especially, to get more accurate values of the moments $\langle\xi^n\rangle_{2;\pi}|_{\mu}$, we will recalculate those moments with the QCD sum rules in the framework of the background field theory (BFTSR) and adopt a more reasonable and accurate sum rules formula for $\langle\xi^n\rangle_{2;\pi}|_{\mu}$.

The remaining parts of this paper are organized as follows. In Sec.~\ref{Sec.III}, we recalculate the pionic leading-twist DA moments by BFTSR. In Sec.~\ref{Sec.II}, we give a brief overview of the LCHO model, put forward the improved new model, and introduce the method of least squares fitting moment to get model parameters. Numerical results are given in Sec.~\ref{Sec.IV}. Section~\ref{Sec.V} is reserved for a summary.

\section{Theoretical framework}

\subsection{The BFTSR for the moments of $\phi_{2;\pi}(x,\mu)$}\label{Sec.III}
To derive the sum rules for the pionic leading-twist DA moments $\langle\xi^n\rangle_{2;\pi}|_{\mu}$, we adopt the following correlation function,
\begin{eqnarray}
\Pi^{(n,0)}_{2;\pi} (z,q) &=& i \int d^4x e^{iq\cdot x} \langle 0| T \{ J_n(x) J^\dag_0(0) \} |0\rangle
\nonumber\\
&=& (z\cdot q)^{n+2} I^{(n,0)}_{2;\pi} (q^2) ,
\label{correlator}
\end{eqnarray}
where $z^2 = 0$, $n=(0,2,4,\cdots)$ since the odd moments vanish due to the isospin symmetry, and the currents
\begin{eqnarray}
J_n(x) &=& \bar{d}(x) {z\!\!\!\slash} \gamma_5 (i z\cdot \tensor{D})^n u(x), \label{CurrentWithVertex} \\
J^\dagger_0(0) &=& \bar{u}(0) {z\!\!\!\slash} \gamma_5 d(0) .
\end{eqnarray}
In physical region, the correlation function (\ref{correlator}) can be calculated by inserting a complete set of intermediate hadronic states. Combining the definition
\begin{align}
\langle 0| \bar{d}(0) {z\!\!\!\slash} \gamma_5 (iz\cdot \tensor{D})^n u(0) |\pi(q)\rangle  = i(z\cdot q)^{n+1} f_\pi \langle\xi^n\rangle_{2;\pi}|_\mu ,
\label{mom}
\end{align}
and the quark-hadron duality, the hadron expression of (\ref{correlator}) can be obtained as
\begin{eqnarray}
\textrm{Im} I^{(n,0)}_{{2;\pi};{\rm had}}(q^2) & =& \pi \delta (q^2 - m_\pi^2) f_\pi^2 \langle\xi^n\rangle_{2;\pi}|_\mu  \nonumber\\
&+& \pi \frac{3}{4\pi^2 (n+1) (n+3)} \theta (q^2 - s_\pi), \label{hadim}
\end{eqnarray}
where $m_\pi$ is the pion mass, $f_\pi$ is the decay constant, $s_\pi$ stands for the continuum threshold. In Eq.~(\ref{mom}), the moments $\langle\xi^n\rangle_{2;\pi}  |_{\mu}$ are defined with the pionic leading-twist DA $\phi_{2;\pi}(x,\mu)$ as following:
\begin{eqnarray}
\langle\xi^n\rangle_{2;\pi}  |_{\mu} = \int^1_0 dx (2x-1)^n \phi_{2;\pi}(x,\mu).
\label{moment}
\end{eqnarray}

In the deep Euclidean region, we apply the operator product expansion (OPE) for the correlation function Eq. (\ref{correlator}). The corresponding calculation is performed in the framework of BFTSR. For the basic assumption of BFTSR, the corresponding Feynman rules, and the OPE calculation technology, one can find in Refs.~\cite{Zhong:2014jla,Huang:1989gv} for detailed discussion.

The hadron expression of correlation function (\ref{correlator}) in the physical region and its OPE in deep Euclidean region can be matched with the dispersion relation. After applying the Borel transformation for both sides, the sum rules for the moments of the pionic leading-twist DA $\phi_{2;\pi}(x,\mu)$ can be obtained as:
\begin{widetext}
\begin{align}
&\frac{\langle\xi^n\rangle_{2;\pi}|_\mu \langle \xi^0\rangle_{2;\pi}|_\mu f_{\pi}^2}{M^2 e^{m_\pi^2/M^2}} = \frac{3}{4\pi^2} \frac{1}{(n+1)(n+3)}  \Big( 1 - e^{-s_\pi/M^2} \Big) + \frac{(m_d + m_u) \langle\bar{q}q \rangle }{(M^2)^2}
~+~ \frac{\langle \alpha_sG^2\rangle }{(M^2)^2}~ \frac{1 + n\theta(n-2)}{12\pi(n+1)}
\nonumber\\[1ex]
&\qquad\qquad~- \frac{(m_d + m_u)\langle  g_s\bar{q}\sigma TGq \rangle }{(M^2)^3}\frac{8n+1}{18} + \frac{\langle g_s\bar{q}q\rangle ^2}{(M^2)^3} \frac{4(2n+1)}{81} - \frac{\langle g_s^3fG^3\rangle }{(M^2)^3}\frac{n \theta(n-2)}{48\pi^2} + \frac{\langle g_s^2\bar{q}q\rangle ^2}{(M^2)^3} \frac{2+\kappa^2}{486\pi^2}
\nonumber\\[1ex]
&\qquad\qquad~ \times \Big\{\!-2(51n+ 25)\Big(\!-\ln \frac{M^2}{\mu^2} \Big) + 3(17n+35) + \theta(n-2)\Big[ 2n  \Big(\!-\ln \frac{M^2}{\mu^2} \Big) + \frac{49n^2 +100n+56}n
\nonumber\\
&\qquad\qquad~ - 25(2n+1)\Big[ \psi\Big(\frac{n+1}{2}\Big) - \psi\Big(\frac{n}{2}\Big) + \ln4 \Big]   \Big] \Big\}.
\label{xinxi0}
\end{align}
Where $M$ is the Borel parameter, and for those vacuum condensates, we have taken:
\begin{eqnarray}
\langle  \bar{q}q \rangle  &=& \langle  \bar{d}d \rangle  = \langle  \bar{u}u \rangle , \nonumber\\
\langle  g_s\bar{q}\sigma TGq \rangle  &=& \langle  g_s\bar{d}\sigma TGd \rangle  = \langle  g_s\bar{u}\sigma TGu \rangle , \nonumber\\
\langle g_s\bar{q}q\rangle ^2 &=& \langle g_s\bar{d}d\rangle ^2 = \langle g_s\bar{u}u\rangle ^2, \nonumber\\
\langle g_s^2\bar{q}q\rangle ^2 &=& \langle g_s^2\bar{d}d\rangle ^2 = \langle g_s^2\bar{u}u\rangle ^2, \nonumber
\end{eqnarray}
and with $\langle  \bar{s}s \rangle /\langle  \bar{q}q \rangle  = \kappa$,
\begin{eqnarray}
g_s^2 \sum\langle g_s\bar{\psi}\psi\rangle ^2 = (2 + \kappa^2) \langle g_s^2\bar{q}q\rangle ^2,~ (\psi = u, d, s). \nonumber
\end{eqnarray}
In the OPE calculation for the correlation function (\ref{correlator}), we have corrected the mistake of a vacuum matrix element, $\langle 0| G^A_{\mu\nu} G^B_{\rho\sigma;\lambda\tau} |0\rangle$, used in the previous work~\cite{Zhong:2014jla}. That is
\begin{align}
\langle 0|G_{\mu \nu }^A G_{\rho \sigma;\lambda \tau }^B|0\rangle  &= {\delta^{AB}}\Big\{ \Big( - \frac1{1296}\sum {{\langle {g_s}\bar \psi \psi \rangle }^2} - \frac1{384}\langle g_s f G^3\rangle\Big) \big[2{g_{\lambda \tau }}({g_{\mu \sigma }}{g_{\nu \rho }} - {g_{\mu \rho }}{g_{\nu \sigma }}) + {g_{\rho \tau }}({g_{\mu \sigma }}{g_{\nu \lambda }}- {g_{\mu \lambda }}{g_{\nu \sigma }})
\nonumber\\
&   + {g_{\sigma \tau }}({g_{\mu \lambda }}{g_{\nu \rho }} - {g_{\mu \rho }}{g_{\nu \lambda }})\big]+ \Big( - \frac1{1296}\sum {{{\langle {g_s}\bar \psi \psi \rangle }^2}}+ \frac1{384}\langle g_s f G^3\rangle \Big) \big[{g_{\mu \tau }}({g_{\rho \nu }}{g_{\sigma \lambda }} - {g_{\rho \lambda }}{g_{\nu \sigma }}) + {g_{\nu \tau }}
\nonumber\\
&\times ({g_{\rho \lambda }}{g_{\sigma \mu }} - {g_{\rho \mu }}{g_{\sigma \lambda }})\big]\Big\}
\label{eq:GG}
\end{align}
\end{widetext}
It needs to be noted that, by taking $n=0$ in Eq.~(\ref{moment}) and considering the normalization of the pionic leading-twist DA $\phi_{2;\pi}(x,\mu)$, one can obtain the $0^{\rm th}$ moment
\begin{eqnarray}
\langle\xi^0\rangle_{2;\pi}  |_{\mu} = 1.
\label{moment0}
\end{eqnarray}
Therefore, in many QCD sum rules calculation people usually substitute Eq.~(\ref{moment0}) as input directly into sum rules (\ref{xinxi0}), and take Eq.~(\ref{xinxi0}) as the sum rules of the moments $\langle \xi^n \rangle_{2;\pi}|_\mu$. This will bring extra deviation to the predicted values of $\langle \xi^n \rangle_{2;\pi}|_\mu$, the reason is that the $0^{\rm th}$ moment $\langle\xi^0\rangle_{2;\pi}  |_{\mu}$ in the l.h.s. of Eq.~(\ref{xinxi0}) is not strict that in Eq.~(\ref{moment0}). By taking $n=0$ in Eq.~(\ref{xinxi0}), one can obtain the sum rule of $\langle\xi^0\rangle_{2;\pi}  |_{\mu}$,
\begin{align}
&\frac{\langle \xi^0\rangle_{2;\pi} ^2|_\mu f_{\pi}^2}{M^2 e^{m_\pi^2/M^2}}
= \frac{1}{4\pi^2} \Big( 1 + \frac{\alpha_s}{\pi} \Big) \Big( 1 - e^{-s_\pi / M^2} \Big) + (m_d
\nonumber\\
&\qquad + m_u)~\frac{ \langle  \bar{q}q \rangle }{(M^2)^2}+ \frac{\langle \alpha_sG^2\rangle }{(M^2)^2} \frac{1}{12\pi} - \frac{1}{18}(m_d + m_u)
\nonumber\\
&\qquad
\times \frac{ \langle  g_s\bar{q}\sigma TGq \rangle }{(M^2)^3} + \frac{4}{81} \frac{\langle g_s\bar{q}q\rangle ^2}{(M^2)^3}
+ \frac{\langle g_s^2\bar{q}q\rangle ^2}{(M^2)^3} \frac{2+\kappa^2}{486\pi^2}
\nonumber\\
&\qquad \times \Big[\!-50 \Big(\!-\ln \frac{M^2}{\mu^2} \Big) + 105 \Big].
\label{xi0xi0}
\end{align}
Obviously, $\langle \xi^0 \rangle_{2;\pi}|_\mu$ in the l.h.s. of sum rule (\ref{xinxi0}) can not be normalized in the whole Borel parameter regions. The reason is that our calculation is not complete. The high-order corrections and high-dimensional corrections have not been calculated, and which are also impossible to calculate completely. In fact, the authors of Ref.~\cite{Xiang:1984hw} discovered this more than $30$ years ago. They obtain $\langle \xi^0 \rangle_{2;\pi}|_\mu \simeq 0.83$, and take $f_\pi \langle \xi^0 \rangle_{2;\pi}|_\mu$ as normalization factor to calculate the values of $\langle \xi^2 \rangle_{2;\pi}|_\mu$ and $\langle \xi^4 \rangle_{2;\pi}|_\mu$. In this paper, we argue that we need to further consider the impact of the sum rule of $\langle \xi^0 \rangle_{2;\pi}|_\mu$, Eq. (\ref{xi0xi0}), in the full Borel parameter regions when using sum rule (\ref{xinxi0}) to calculate $\langle \xi^n \rangle_{2;\pi}|_\mu$. Therefore, in order to obtain more accurate moments $\langle \xi^n \rangle_{2;\pi}|_\mu$, we suggest the following form:
\begin{eqnarray}
\langle\xi^n\rangle_{2;\pi}|_\mu  = \frac{(\langle\xi^n\rangle_{2;\pi}|_\mu \langle \xi^0\rangle_{2;\pi}|_\mu)|_{\rm From\ Eq.~(7)}}{\sqrt{\langle \xi^0\rangle_{2;\pi}^2|_\mu} |_{\rm From\ Eq.~(10)}}.
\label{xin}
\end{eqnarray}
Meanwhile, another advantage of Eq.~(\ref{xin}) is that it can also eliminate some systematic errors caused by the continuum state, the absence of high dimensional condensates, the selection and determination of various input parameters.

It should be mentioned that Eq. (\ref{xi0xi0}) is usually used to predict the pion decay constant $f_\pi$ on the premise that $\langle\xi^0\rangle_{2;\pi}|_\mu \equiv 1$. After the previous discussion, we think that the sum rule of $\langle\xi^0\rangle_{2;\pi}|_\mu$ varies with Borel parameter $M^2$, especially when the $f_\pi$ has a definite experimental value. In order to ensure QCD sum rule's prediction ability on other meson decay constant, we need to assume that $\langle\xi^0\rangle_{2;\pi}|_\mu$ can be normalized in a appropriate Borel window.

\subsection{The improved LCHO model for $\phi_{2;\pi}(x,\mu)$}\label{Sec.II}
\begin{table*}
\caption{The expressions of the spin-space wave function $\chi_{2;\pi}^{\lambda_1 \lambda_2}(x,\textbf{k}_\perp)$ with different $\lambda_1 \lambda_2$.}\label{table:chi}
\begin{tabular}{c c c c c}
\hline\hline
~~~$\lambda_1 \lambda_2$~~~ & ~~~$\downarrow\downarrow$~~~ & ~~~$\uparrow\uparrow$~~~ & ~~~$\uparrow\downarrow$~~~ & ~~~$\downarrow\uparrow$
\\ \hline
~~~$\chi_{2;\pi}^{\lambda_1 \lambda_2}(x,\textbf{k}_\perp)$~~~ & ~~~$-\dfrac{k_x + i k_y}{\sqrt{2(m_q^2 + \textbf{k}_\perp^2)}}$~~~ & ~~~$-\dfrac{k_x - i k_y}{\sqrt{2(m_q^2 + \textbf{k}_\perp^2)}}$~~~ & ~~~$ \dfrac{m_q}{\sqrt{2(m_q^2 + \textbf{k}_\perp^2)}}$~~~ & ~~~$ - \dfrac{m_q}{\sqrt{2(m_q^2 + \textbf{k}_\perp^2)}}$  \\ \hline\hline
\end{tabular}
\end{table*}
Based on the BHL-description~\cite{BHL}, the LCHO model of the pion leading-twist WF has raised in Refs.~\cite{Wu:2010zc, Wu:2011gf}, and its form is:
\begin{eqnarray}
\Psi_{2;\pi}(x,\textbf{k}_\bot) = \sum_{\lambda_1\lambda_2} \chi_{2;\pi}^{\lambda_1\lambda_2}(x,\textbf{k}_\bot) \Psi^R_{2;\pi}(x,\textbf{k}_\bot),
\label{WF_full}
\end{eqnarray}
where $\textbf{k}_\bot$ is the pionic transverse momentum, $\lambda_1$ and $\lambda_2$ are the helicities of the two constituent quark. $\chi_{2;\pi}^{\lambda_1\lambda_2}(x,\textbf{k}_\bot)$ stands for the spin-space WF that comes from the Wigner-Melosh rotation, whose explicit form for different $\lambda_1\lambda_2$ are exhibited in Table \ref{table:chi}, which can also been seen in Refs.~\cite{Huang:1994dy,Cao:1997hw,Huang:2004su,Wu:2005kq}.
\begin{eqnarray}
\Psi^R_{2;\pi}(x,\textbf{k}_\bot) = A_{2;\pi} \varphi_{2;\pi}(x) \exp \left[ -\frac{\textbf{k}^2_\bot + m_q^2}{8\beta_{2;\pi}^2 x\bar x} \right],
\label{WF_spatial}
\end{eqnarray}
indicates spatial WF, where $\bar{x} = 1-x$, $A_{2;\pi}$ is the normalization constant, the $\textbf{k}_\bot$-dependence part of the spatial WF $\Psi^R_{2;\pi}(x,\textbf{k}_\bot)$ comes from the approximate bound-state solution in the quark model for pion~\cite{WF_restframe} and determine the WF's transverse distribution via the harmonious parameter $\beta_{2;\pi}$, the $x$-dependence part $\varphi_{2;\pi}(x)$ dominates the WF's longitudinal distribution. In principle, the spatial WF $\Psi^R_{2;\pi}(x,\textbf{k}_\bot)$ should include a Jacobi factor. Numerical calculation in Sec. IIIC will show that the Jacobi factor has a little effect on the behavior of the the pionic leading-twist DA. In Table \ref{table:chi} and Eq.~(\ref{WF_spatial}), $m_q$ stands for the mass of the constitute quark $u$ and $d$ in pion. In our previous work~\cite{Zhong:2015nxa}, the experimental data of the pion-photon transition form factor reported by the CELLO, CLEO, BABAR and BELLE collaborators based on the LCHO model with the longitudinal distribution function $\varphi^{\rm I}_{2;\pi}(x)$ (see Eq.~(\ref{varphiI})) have been fit by adopt the least square method. Where we take the constituent quark mass $m_q$ and the model parameter $B$ as the fitting parameters, and obtain $m_q = (216, 246, 347, 222)~{\rm MeV}$ for CELLO, CLEO, BABAR and BELLE data, respectively. The corresponding goodness-of-fit are $P_{\chi^2_{\rm min}}/n_d = (0.187/3, 0.986/13, 0.416/15, 0.958/13)$. Then we take $m_q =200~{\rm MeV}$ in this paper. Otherwise, $m_q$ is taken to be $250 \rm MeV$ in the invariant meson mass scheme~\cite{Terentev:1976jk, Jaus:1989au, Jaus:1991cy, Chung:1988mu,Choi:1997qh,Schlumpf:1994bc, Cardarelli:1994yq} and $330 \rm MeV$ in the spin-averaged meson mass scheme~\cite{Dziembowski:1986dr,Dziembowski:1987zp,Ji:1990rd,Ji:1992yf,Choi:1996mq}. Therefore, we will discuss the impact of different values of $m_q$ on the behavior of our pionic leading-twist DA in detail by taking $m_q=200 \sim 350~{\rm MeV}$.

Using the relationship between the pionic leading-twist DA and WF,
\begin{eqnarray}
\phi_{2;\pi}(x,\mu) = \frac{2\sqrt{6}}{f_\pi} \int_{| \mathbf{k}_\bot |^2 \leq \mu^2} \frac{d^2\mathbf{k}_\bot}{16\pi^3} \Psi_{2;\pi}(x,\mathbf{k}_\bot),
\label{DA_WF}
\end{eqnarray}
the leading-twist DA for pion, $\phi_{2;\pi}(x,\mu)$, can be obtained. That is, after integrating over the transverse momentum $\mathbf{k}_\bot$ in Eq.~(\ref{DA_WF}), we have
\begin{eqnarray}
&&\phi_{2;\pi}(x,\mu) = \frac{\sqrt{3} A_{2;\pi} m_q \beta_{2;\pi}}{2\pi^{3/2}f_\pi} \sqrt{x\bar x} \varphi_{2;\pi}(x) \nonumber\\
&&\qquad \times \left\{ \textrm{Erf}\left[ \sqrt{\frac{m_q^2 + \mu^2}{8\beta_{2;\pi}^2 x\bar x}} \right]
-  \textrm{Erf}\left[ \sqrt{\frac{m_q^2}{8\beta_{2;\pi}^2 x\bar x}} \right] \right\},\nonumber\\
\label{DA_model}
\end{eqnarray}
where ${\rm Erf}(x) = 2\int^x_0 e^{-t^2} dx/{\sqrt{\pi}}$ is the error function. The error function part in Eq.~(\ref{DA_model}) comes from the $\textbf{k}_\bot$-dependence part of the WF $\Psi_{2;\pi}(x,\textbf{k}_\bot)$ and gives a good endpoint behavior for $\phi_{2;\pi}(x,\mu)$, and $\varphi_{2;\pi}(x)$ dominates the broadness of $\phi_{2;\pi}(x,\mu)$. Obviously, the specific form of $\phi_{2;\pi}(x,\mu)$ is determined by the parameters $A_{2;\pi}$, $\beta_{2;\pi}$ and the function $\varphi_{2;\pi}(x)$. There are two important constraints~\cite{BHL} which can be used to constrain the parameters $A_{2;\pi}$ and $\beta_{2;\pi}$, that is,
\begin{itemize}
  \item[(1)] the WF normalization condition provided from the process $\pi \to \mu\nu$,
      \begin{eqnarray}
      \int^1_0 dx \int \frac{d^2 \textbf{k}_\bot}{16\pi^3} \Psi (x,\textbf{k}_\bot) = \frac{f_\pi}{2\sqrt{6}};
      \label{DA_constraint1}
      \end{eqnarray}
  \item[(2)] the sum rule derived from $\pi^0 \to \gamma\gamma$ decay amplitude,
      \begin{eqnarray}
      \int^1_0 dx \Psi(x, \textbf{k}_\bot = \textbf{0}) = \frac{\sqrt{6}}{f_\pi}.
      \label{DA_constraint2}
      \end{eqnarray}
\end{itemize}

Then the pionic leading-twist DA $\phi_{2;\pi}(x,\mu)$ only depends on the mathematical form of $\varphi_{2;\pi}(x)$. By solving the renormalization group equation of the pionic leading-twist DA, $\phi_{2;\pi}(x,\mu)$ can be written as the expansion form of the Gegenbauer series~\cite{Lepage:1979zb, Efremov:1979qk}. Based on this, in our previous paper, $\varphi_{2;\pi}(x)$ is taken to be the linear superposition of the first several Gegenbauer polynomials. For example, in Refs.~\cite{Wu:2012kw, Huang:2013gra, Huang:2013yya, Zhong:2015nxa}, we take
\begin{eqnarray}
\varphi^{\rm I}_{2;\pi}(x) = 1 + B\times C^{3/2}_2(2x-1),
\label{varphiI}
\end{eqnarray}
and
\begin{eqnarray}
\varphi^{\rm II}_{2;\pi}(x) &=& 1 + B_2\times C^{3/2}_2(2x-1) \nonumber\\
&+& B_4\times C^{3/2}_4(2x-1)
\label{varphiII}
\end{eqnarray}
is adopted in Ref.~\cite{Zhong:2014jla}. For the former, when the value of parameter $B$ changes from $0.0$ to $0.6$, the pionic leading-twist DA model, i.e. Eq.~(\ref{DA_model}) can mimic the DA behavior from asymptotic-like to CZ-like. And for the latter, we further consider the correction of $4^{\rm th}$ order Gegenbauer polynomial. The mathematical form of $\varphi_{2;\pi}(x)$ can usually be determined in two ways. The first one is to extract $\varphi_{2;\pi}(x)$ from the experimental data of the exclusive processes involving pion~\cite{Wu:2012kw, Huang:2013gra, Huang:2013yya, Zhong:2015nxa}, such as semi-leptonic decays $B\to \pi \ell\nu_\ell$ and $D\to \pi \ell\nu_\ell$, the pion-photon transition form factor $F_{\pi\gamma}(Q^2)$, and the exclusive process $B^0\to\pi^0\pi^0$, etc.; the second one is to determine from the moments $\langle\xi^n\rangle_{2;\pi}|_{\mu}$ or the Gegenbauer moments $a^{2;\pi}_n(\mu)$ of $\phi_{2;\pi}(x,\mu)$. In Ref.~\cite{Zhong:2014jla}, we have adopted the second method to determine the mathematical form of $\varphi_{2;\pi}(x)$ and further the behavior of $\phi_{2;\pi}(x,\mu)$.

In this paper, we will still make use of the second method mentioned above to determine the behavior of $\phi_{2;\pi}(x,\mu)$, but we will improve it. The accuracy of the behavior of $\phi_{2;\pi}(x,\mu)$ obtained by this method is restricted by two aspects: the rationality of the constructed mathematical form of $\varphi_{2;\pi}(x)$ and the accuracy of moments. In order to get better mathematical form of $\varphi_{2;\pi}(x)$, a natural idea is to add higher order Gegenbauer polynomial correction in $\varphi^{\rm II}_{2;\pi}(x)$, as we have done for the $D,\eta_c,B_c, \eta_b$ twist-2, 3 DAs in Refs.~\cite{Zhong:2014fma, Zhong:2016kuv, Zhang:2017rwz, Zhong:2018exo}. However, such improvement obviously destroys the beauty and conciseness of the model. Otherwise, we find that the parameters $B_2,\ B_4$ are close to the Gegenbauer moments $a_2^{2;\pi}(\mu),\ a_4^{2;\pi}(\mu)$ respectively. From the relationship between $\langle\xi^n\rangle_{2;\pi}|_{\mu}$ and $a_n^{2;\pi}(\mu)$, it can be seen that the reliability of $a_n^{2;\pi}(\mu)$ calculated by QCD sum rules decreases sharply with the increase of order-$n$. In view of this, in this paper we will improve the mathematical form of $\varphi_{2;\pi}(x)$ by other way, as well as propose a new determination method of model parameters.

We notice that although it is difficult to improve pionic leading-twist DA by introducing higher Gegenbauer polynomial correction, our goal is still to make it more reasonable and accurate by adjusting the behavior of $\phi_{2;\pi}(x,\mu)$. We find that the factor $\sqrt{x\bar x}$ in Eq.~(\ref{DA_model}) can regulate DA's behavior to some extent. Inspired by this, we introduce a factor $\left[x\bar x\right]^{\alpha_{2;\pi}}$ into WF's longitudinal distribution function $\varphi_{2;\pi}(x)$, i.e.,
\begin{align}
\varphi^{\rm III}_{2;\pi}(x) = \left[ x\bar x \right]^{\alpha_{2;\pi}}.
\label{varphiIII}
\end{align}
In order to further apply our LCHO model to other meson DA, and combine the form of $\varphi^{\rm I}_{2;\pi}(x)$, we propose a more complex form,
\begin{align}
\varphi^{\rm IV}_{2;\pi}(x) = \left[ x\bar x \right]^{\alpha_{2;\pi}} \left[ 1 + \hat{a}^{2;\pi}_2 C_2^{3/2}(2x-1) \right],
\label{varphiIV}
\end{align}
the parameters $\alpha_{2;\pi}$ and $\hat{a}^{2;\pi}_2$ will be determined by fitting the moments $\langle\xi^n\rangle_{2;\pi}|_{\mu}$ directly through the method of least squares, and the values of moments $\langle\xi^n\rangle_{2;\pi}|_{\mu}$ come from Eq.~(\ref{xin}) calculated under BFTSR in Sec.~\ref{Sec.III}. In order to distinguish our LCHO model with $\varphi^{\rm III}_{2;\pi}(x)$ and $\varphi^{\rm IV}_{2;\pi}(x)$, and facilitate the discussion later, we will record the former as LCHO model-III and the latter as LCHO model-IV.

Considering a set of $N$ independent measurements $y_i$ with the known variance $\sigma_i$ and the mean $\mu(x_i;\mathbf{\theta})$ at known points $x_i$. The objective of the least squares method is to obtain the best value of fitting parameters $\mathbf{\theta}$ by minimizing the likelihood function~\cite{PDGnew}
\begin{eqnarray}
\chi^2(\mathbf{\theta}) =
\sum^N_{i=1} \frac{(y_i - \mu(x_i,\mathbf{\theta}))^2}{\sigma_i^2}.
\label{ls}
\end{eqnarray}
As for the present case, the function $\mu(x_i;\mathbf{\theta})$ indicates the pionic leading-twist DA moments $\langle\xi^n\rangle_{2;\pi}|_{\mu}$ defined by combining Eqs.~\eqref{moment}, \eqref{DA_model}, \eqref{varphiIII} and $\mathbf{\theta} = (\alpha_{2;\pi}, \hat{a}^{2;\pi}_2)$; The theoretical values of $\langle\xi^n\rangle_{2;\pi}|_{\mu}$ calculated by QCD sum rules in next section are assumed to be the value of $y_i$ and its variance $\sigma_i$. The probability density function of $\chi^2$ can be obtained,
\begin{eqnarray}
f(y;n_d) = \frac{1}{\Gamma\left(\dfrac{n_d}{2}\right) 2^{n_d/2}} y^{\frac{n_d}{2}-1} e^{-\frac{y}{2}},
\end{eqnarray}
$n_d$ is the number of degree-of-freedom. Then one can further calculate the following probability,
\begin{eqnarray}
P_{\chi^2} = \int^\infty_{\chi^2} f(y;n_d) dy.
\label{px2}
\end{eqnarray}
The magnitude of the probability $P_{\chi^2}$ ($P_{\chi^2} \in [0,1]$) can be used to judge the goodness-of-fit, when its value is closer to $1$, a better fit is assumed to be achieved.

\section{numerical analysis}\label{Sec.IV}

\subsection{Basic input parameters}
To do the numerical calculation, we adopt the latest data from Particle Data Group (PDG)~\cite{PDGnew}: $m_\pi = 139.57039 \pm 0.00017~\textrm{MeV}$ and $f_\pi = 130.2\pm1.2~\textrm{MeV}$. The current-quark-mass for the $u, d$-quark are adopted as $m_u = 2.16^{+0.49}_{-0.26}~\textrm{MeV}$ and $m_d = 4.67^{+0.48}_{-0.17}~\textrm{MeV}$ at scale $\mu = 2 ~\rm GeV$. Based on these latest values, we can update the vacuum condensates.
\begin{itemize}
\item For the double-quark condensate, we adopt Gell-Mann-Oakes-Renner relation:
\begin{eqnarray}
&&m_u \langle \bar{u}u\rangle  + m_d \langle \bar{d}d\rangle  \simeq - \frac{ f_\pi^2 m_\pi^2 }{2}
\nonumber\\
&&\qquad= -(1.651 \pm 0.003) \times 10^{-4}~\textrm{GeV}^4.
\label{mqq}
\end{eqnarray}
Combining with the $u,d$ quark masses, we have:
\begin{eqnarray}
\langle \bar{q}q\rangle  &=& \left( -2.417_{-0.114}^{+0.227} \right) \times 10^{-2}~{\rm GeV}^3 \nonumber\\
&=& \left( -289.14_{-4.47}^{+9.34} \right)^3 ~{\rm MeV}^3,
\label{qq}
\end{eqnarray}
at scale $\mu = 2~\rm GeV$.

\item By combining Eqs.~\eqref{mqq}, \eqref{qq} and the relation $\langle g_s\bar{q}\sigma TGq\rangle  = m_0^2 \langle \bar{q}q\rangle$ with $m_0^2 = 0.80 \pm 0.02~\textrm{GeV}^2$~\cite{Narison:2014ska}, the quark-gluon mixed condensate would be
\begin{eqnarray}
&& m_u \langle g_s\bar{u}\sigma TGu\rangle  + m_d \langle g_s\bar{d}\sigma TGd\rangle
\nonumber\\
&& \quad\quad\quad = -(1.321 \pm 0.033) \times 10^{-4}~{\rm GeV}^6 , \label{mqgq}
\end{eqnarray}
\begin{eqnarray}
\langle g_s\bar{q}\sigma TGq\rangle  &=& \left( - 1.934^{+0.188}_{-0.103} \right) \times 10^{-2}~{\rm GeV}^5.
\label{qgq}
\end{eqnarray}

\item By adopting the data in Ref.~\cite{Narison:2014ska},
\begin{eqnarray}
\rho \alpha_s \langle \bar{q}q\rangle^2  = \left( 5.8 \pm 1.8 \right) \times 10^{-4}~{\rm GeV}^6, \label{rhogq24}
\end{eqnarray}
with $\rho \simeq 3 - 4$, and combining the value of the double-quark condensate in Eq.~(\ref{qq}), the four-quark condensates can be obtained as:
\begin{eqnarray}
\langle g_s\bar{q}q\rangle ^2 = (2.082^{+0.734}_{-0.697}) \times 10^{-3} ~\textrm{GeV}^6,
\label{gq24}
\end{eqnarray}
and
\begin{eqnarray}
\langle g_s^2\bar{q}q\rangle ^2 = (7.420^{+2.614}_{-2.483}) \times 10^{-3} ~\textrm{GeV}^6,
\label{gq44}
\end{eqnarray}

\item From Ref.~\cite{Colangelo:2000dp}, we have
\begin{equation}
\langle \alpha_s G^2\rangle  = 0.038 \pm 0.011 ~\textrm{GeV}^4,
\label{gg}
\end{equation}
and
\begin{eqnarray}
\langle g_s^3fG^3\rangle  \simeq 0.045 ~\textrm{GeV}^6.
\label{ggg}
\end{eqnarray}

\item For the ratio $\kappa = \langle  \bar{s}s \rangle /\langle  \bar{q}q \rangle$, Ref.~\cite{Narison:2014wqa} gives:
\begin{eqnarray}
\kappa = 0.74 \pm 0.03,
\label{kappa}
\end{eqnarray}
\end{itemize}

\subsection{The renormalization group equation for the input parameters and the moments $\langle\xi^n\rangle_{2;\pi}|_{\mu}$}

In numerical calculation for the moments' BFTSR (\ref{xin}), we take the scale $\mu = M$ as usual. From $\alpha_s(M_z) = 0.1179 \pm 0.0010$ with $M_Z = 91.1876 \pm 0.0021~ {\rm GeV}$, and combining $\bar{m}_c(\bar{m}_c) = 1.27 \pm 0.02~{\rm GeV}$ and $\bar{m}_b(\bar{m}_b) = 4.18^{+0.03}_{-0.02}~{\rm GeV}$~\cite{PDGnew}, under the 3-loop approximate solution we predict $\Lambda_{\rm QCD}^{(n_f)} \simeq 324, 286, 207~{\rm MeV}$ for the number of quark flavors $n_f = 3, 4, 5$, respectively.

The renormalization group equations (RGE) of the quark mass and vacuum condensates are given as~\cite{Yang:1993bp, Hwang:1994vp, Lu:2006fr}:
\begin{eqnarray}
m_q|_\mu &=& m_q|_{\mu_0} \left[ \frac{\alpha_s(\mu_0)}{\alpha_s(\mu)} \right]^{-4/\beta_0}, \nonumber\\
\langle \bar{q}q\rangle |_\mu &=& \langle \bar{q}q\rangle |_{\mu_0} \left[ \frac{\alpha_s(\mu_0)}{\alpha_s(\mu)} \right]^{4/\beta_0}, \nonumber\\
\langle g_s\bar{q}\sigma TGq\rangle |_\mu &=& \langle g_s\bar{q}\sigma TGq\rangle |_{\mu_0} \left[ \frac{\alpha_s(\mu_0)}{\alpha_s(\mu)} \right]^{-2/(3\beta_0)}, \nonumber\\
\langle \alpha_s G^2\rangle |_\mu &=& \langle \alpha_s G^2\rangle |_{\mu_0}, \nonumber\\
\langle g_s^3fG^3\rangle |_\mu &=& \langle g_s^3fG^3\rangle |_{\mu_0},
\label{RenormalizationGroupEquation}
\end{eqnarray}
with $\beta_0 = (33-2n_f)/3$. Obviously, the double-gluon condensate and the triple-gluon condensate are energy scale independent. From Eq.~(\ref{eq:GG}), one can find that $\langle g_s^2\bar{q}q\rangle ^2$ and $\langle g_s^3 f G^3\rangle$ have the same RGE. In other words, $\langle g_s^2\bar{q}q\rangle^2$ is also energy scale independent, e.g.,
\begin{eqnarray}
\langle g_s^2\bar{q}q\rangle ^2|_\mu = \langle g_s^2\bar{q}q\rangle ^2|_{\mu_0}.
\label{RGEgg24}
\end{eqnarray}
Combining with the RGE of the double-quark condensate and Eq.~(\ref{RGEgg24}), one can find that the $\langle g_s\bar{q}q\rangle ^2$ and $\langle \bar{q}q\rangle$ have the same energy scale evolution equation, e.g.,
\begin{eqnarray}
\langle g_s\bar{q}q\rangle^2|_\mu &=& \langle g_s\bar{q}q\rangle^2|_{\mu_0} \left[ \frac{\alpha_s(\mu_0)}{\alpha_s(\mu)} \right]^{4/\beta_0}.
\end{eqnarray}
It should be noted that, according to the basic assumption of BFTSR, $g_s$ in all the above vacuum condensates is the ``coupling constant'' between the background fields, which is different from the one in pQCD, and should be absorbed into vacuum condensates as part of these non-perturbative parameters.
\begin{figure*}[t]
\centering
\includegraphics[width=0.33\textwidth]{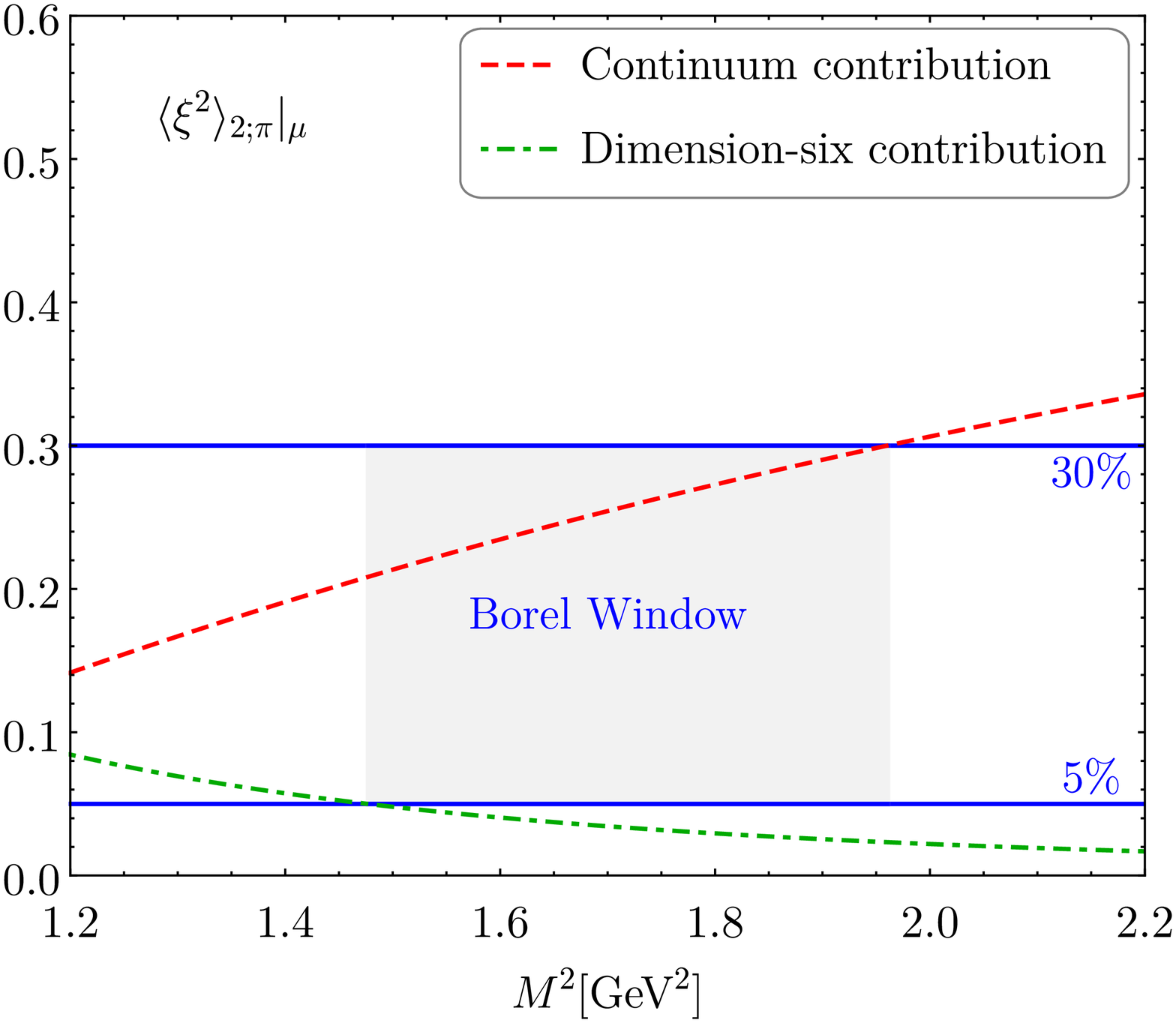}\includegraphics[width=0.33\textwidth]{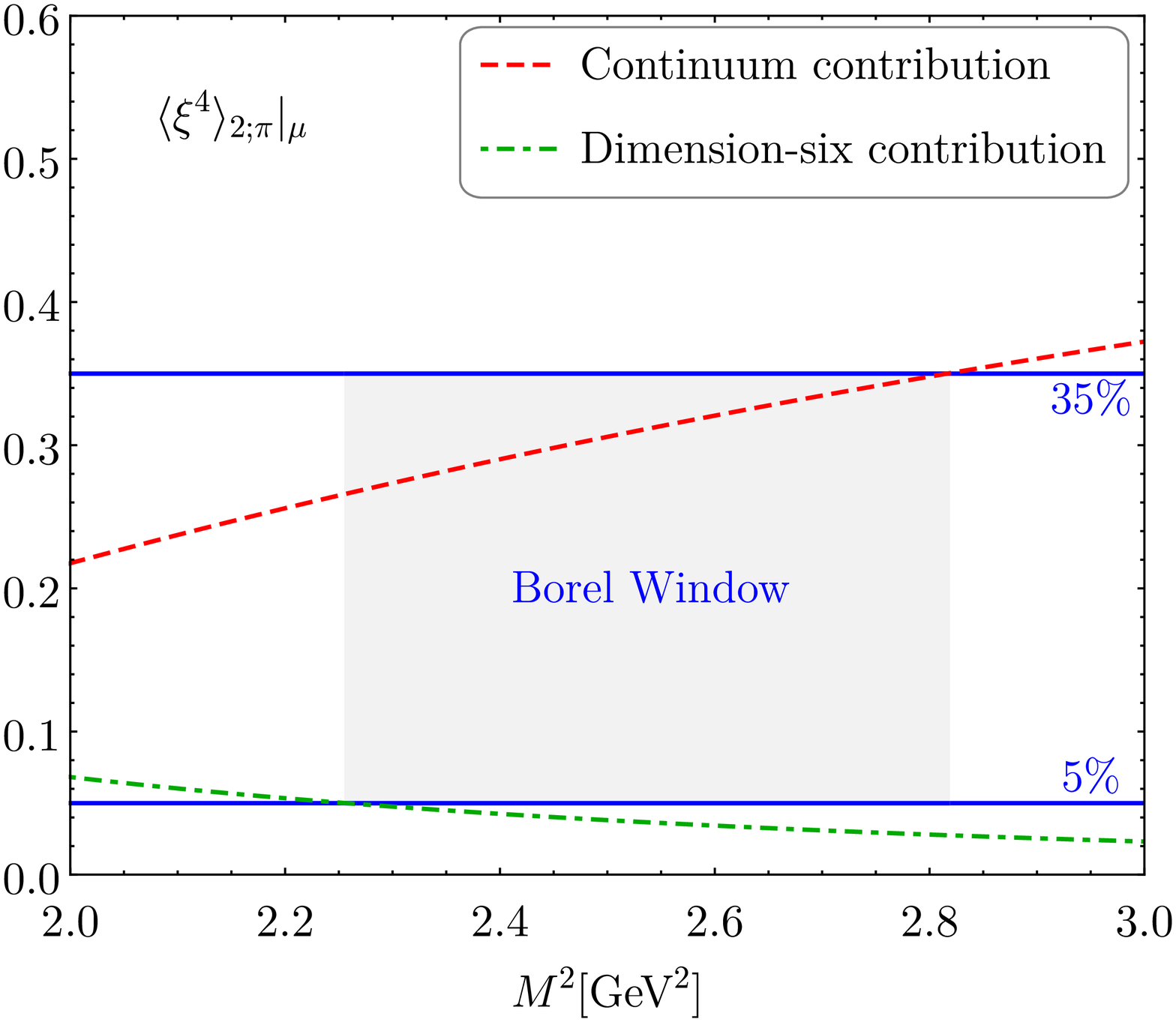}\includegraphics[width=0.33\textwidth]{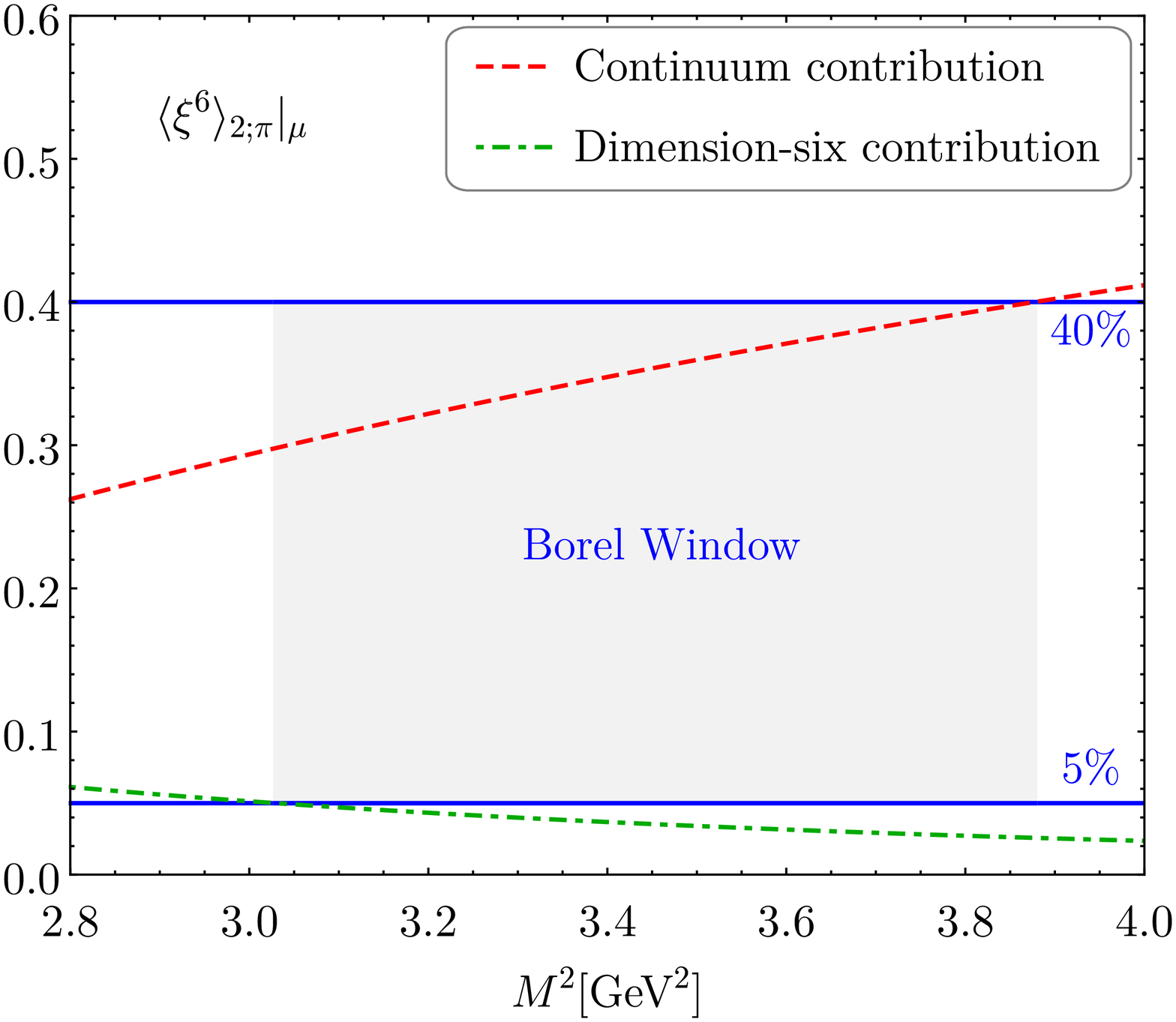}
\includegraphics[width=0.33\textwidth]{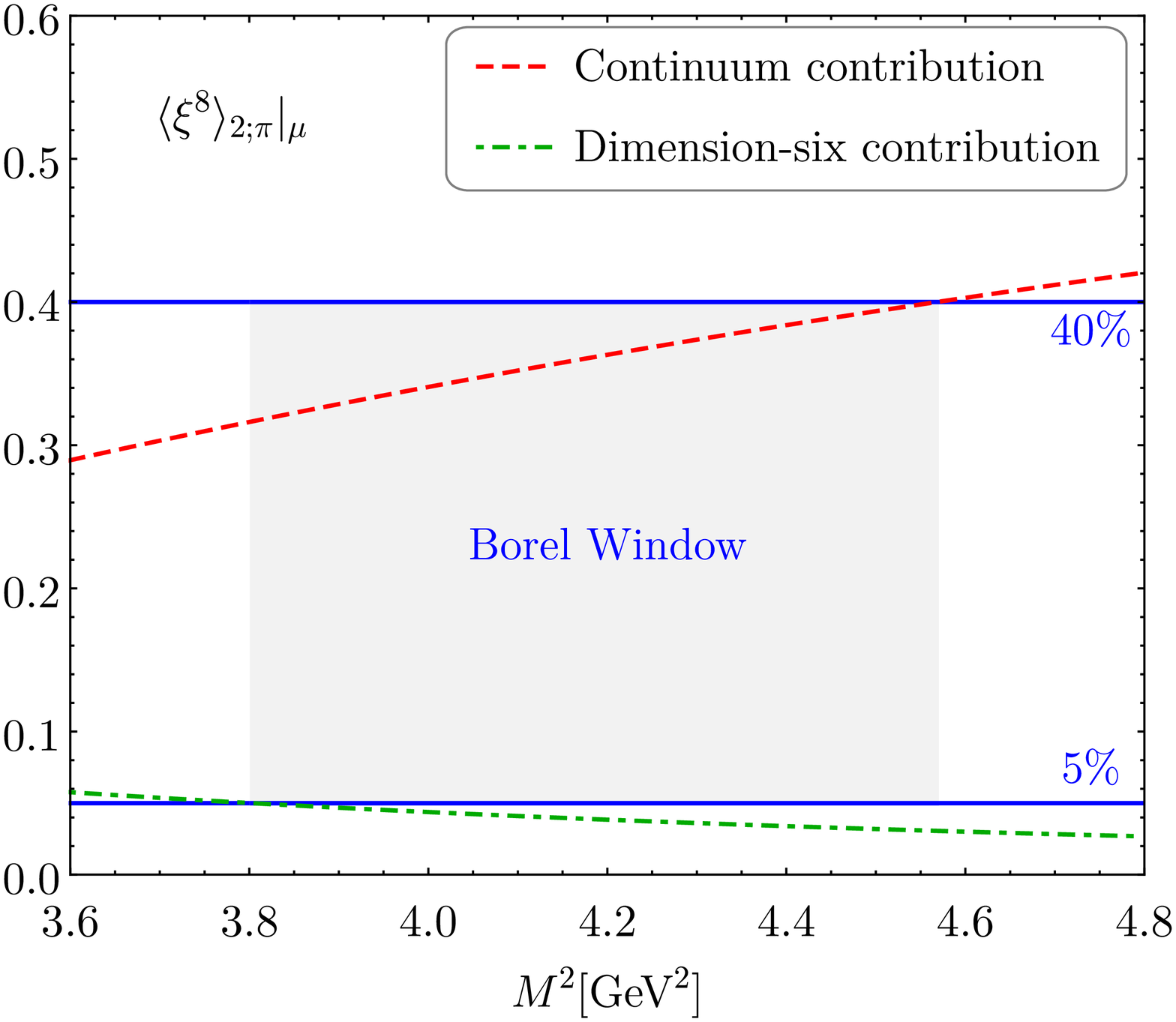}\includegraphics[width=0.325\textwidth]{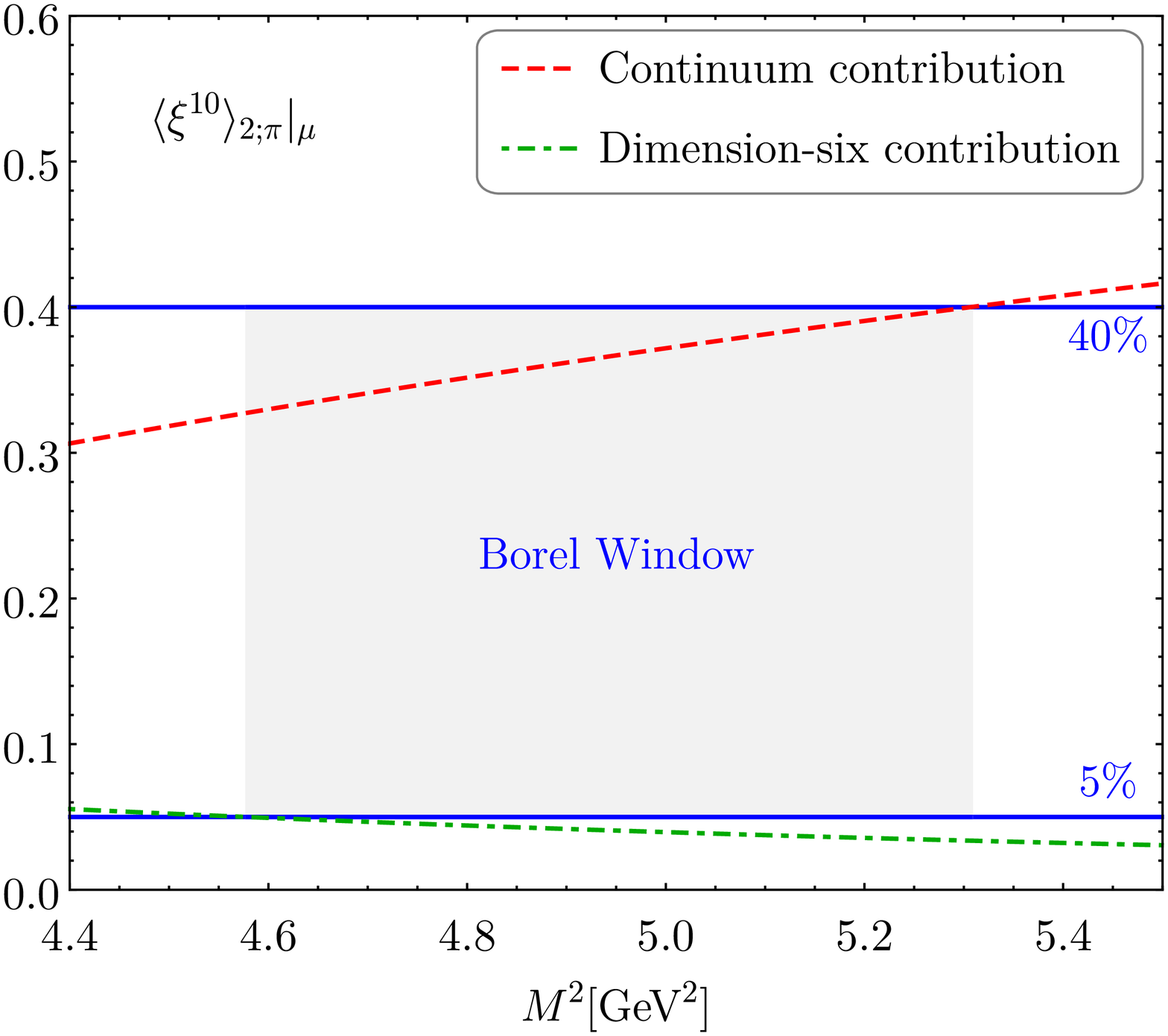}
\caption{The continuum state and dimension-six contribute to the pionic leading-twist DA moments $\langle\xi^n\rangle_{2;\pi}|_\mu$ versus the Borel parameter $M^2$ within BFTSR approach. The shaded band indicate the Borel Window for $\langle\xi^n\rangle_{2;\pi}|_\mu$ for $n=(2,4,6,8,10)$ respectively.}
\label{fRatio}
\end{figure*}

The RGE of the Gegenbauer moments of the pion leading-twist distribution amplitude is:
\begin{eqnarray}
a_n^{2;\pi}(\mu) &=& a_n^{2;\pi}(\mu_0) E_n(\mu, \mu_0),
\label{anRGE}
\end{eqnarray}
with
\begin{eqnarray}
E_n(\mu, \mu_0) &=& \left[ \frac{\alpha_s(\mu)}{\alpha_s(\mu_0)} \right]^{\gamma_n^{(0)}/(2\beta_0)}. \nonumber
\end{eqnarray}
The LO anomalous dimension
\begin{eqnarray}
\gamma_n^{(0)} = 8C_F \left[ \psi(n+2) + \gamma_E - \frac{3}{4} - \frac{1}{2(n+1)(n+2)} \right],
\nonumber
\end{eqnarray}
with $C_F = 4/3$. Based on Eq.~(\ref{anRGE}), the RGE of the moments $\langle\xi^n\rangle_{2;\pi}|_{\mu}$ can be obtained.

With the BFTSR of the moments of the pionic leading-twist distribution amplitude $\phi_{2;\pi}(x,\mu)$ shown in Eqs. (\ref{xinxi0}), (\ref{xi0xi0}) and (\ref{xin}), the values of $\langle\xi^n\rangle_{2;\pi}|_{\mu}$ can be calculated. By requiring that there is reasonable Borel window to normalize $\langle \xi^0\rangle_{2;\pi}|_{\mu}$ with Eq.~(\ref{xi0xi0}), one can get the continuum threshold parameter as about $s_\pi \simeq 1.05 {\rm GeV}^2$. In addition to the traditional method to determine the contribution of the continuum state, the continuum method can limiting or overcoming model-dependence and drawing clean lines in connecting the data with QCD itself~\cite{Qin:2020rad}. To obtain the allowable Borel window for the sum rules of $\langle\xi^n\rangle_{2;\pi}|_{\mu}$, we require that the continuum state's contribution and the dimension-six condensate's contribution to be as small as possible, and the values for $\langle\xi^n\rangle_{2;\pi}|_{\mu}$ are stable in the Borel window. Based on the criteria, the Dimension-six contribution for $\langle\xi^n\rangle_{2;\pi}|_{\mu}$ are prescribe a limit to less than $5\%$ for all the $n$th-order. And the continuum contribution for $\langle\xi^n\rangle_{2;\pi}|_{\mu}$ are restrict to $(30, 35, 40, 40, 40)\%$ for $n=(2,4,6,8,10)$ respectively.

To have a deeper insight into the continuum state and dimension-six contribute to the pionic leading-twist DA moments $\langle\xi^n\rangle_{2;\pi}|_\mu$ versus the Borel parameter $M^2$ within BFTSR approach, we present the curves in Fig.~\ref{fRatio}. The shaded band indicate the Borel Window for $\langle\xi^n\rangle_{2;\pi}|_\mu$ for $n=(2,4,6,8,10)$ respectively. The figure indicates that,
\begin{itemize}
  \item The dimension-six contributions are constraint in the region $<5\%$ guaranteed good convergence for the BFTSR results. And the continuum contributions no more than $40\%$ have agreement with the traditional sum rule strictly.
  \item Borel parameters associate with the region of Borel Window become larger with the increase of index $n$.
\end{itemize}

\begin{table}[t]
\caption{The determined Borel windows and the corresponding pionic leading-twist DA moments $\langle\xi^n\rangle_{2;\pi}|_{\mu}$ with $n=(2,4,6,8,10)$. Where all input parameters are set to be their central values.}
\begin{tabular}{ l  c  c }
\hline\hline
$n$ & ~~~~~~~~~~~~~~~~~~~~$M^2$~~~~~~~~~~~~~~~~~~~~ & $\langle\xi^n\rangle_{2;\pi}|_{\mu}$ \\
\hline
$2$~ & ~$[1.477,1.961]$~ & ~$[0.268,0.257]$~ \\
$4$~ & ~$[2.257,2.817]$~ & ~$[0.132,0.124]$~ \\
$6$~ & ~$[3.029,3.878]$~ & ~$[0.081,0.074]$~ \\
$8$~ & ~$[3.803,4.568]$~ & ~$[0.057,0.052]$~ \\
$10$~ & ~$[4.579,5.307]$~ & ~$[0.042,0.039]$~ \\
\hline\hline
\end{tabular}
\label{tbw}
\end{table}
\begin{figure}[t]
\centering
\includegraphics[width=0.42\textwidth]{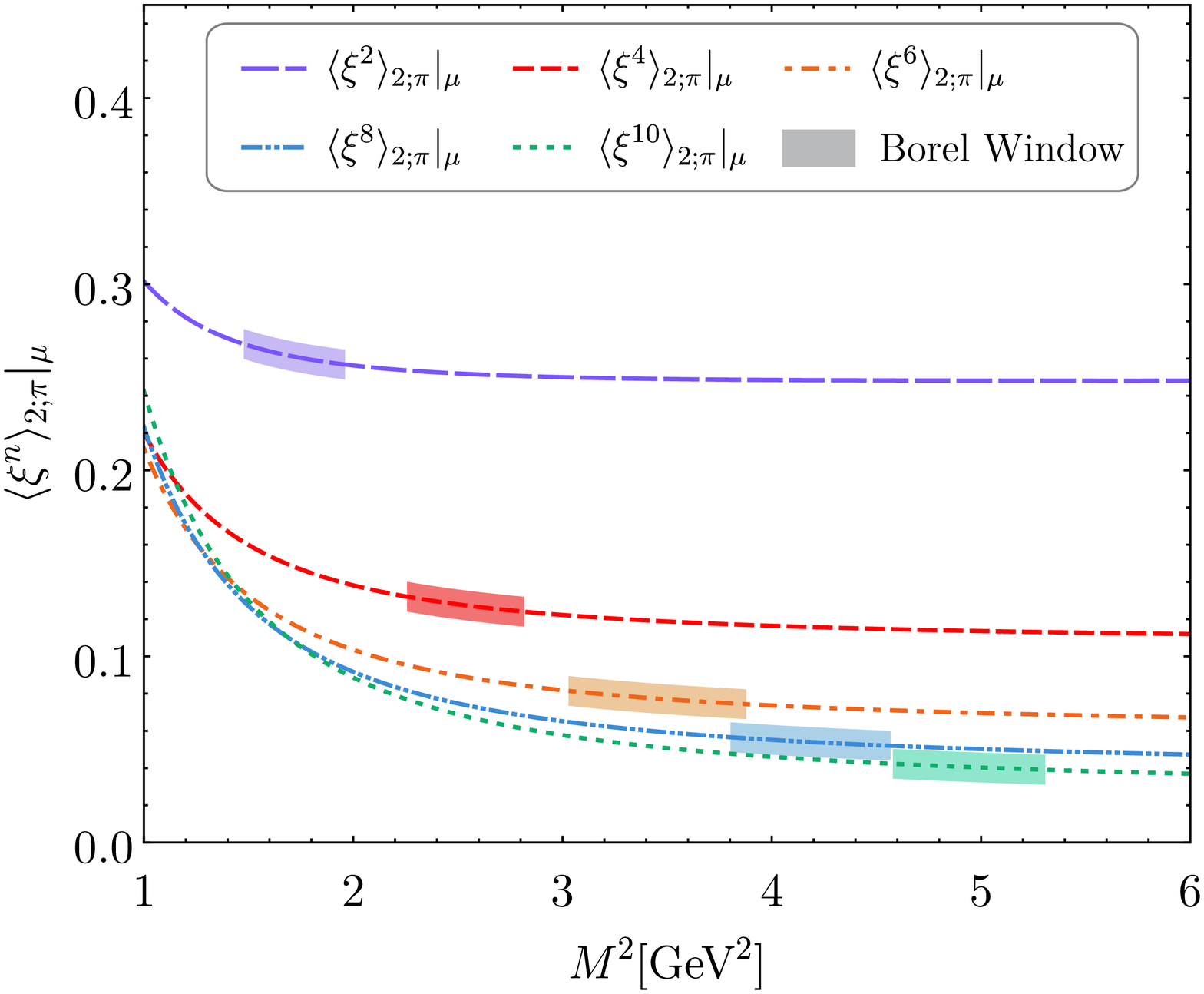}
\caption{The pionic leading-twist DA moments $\langle\xi^n\rangle_{2;\pi}|_{\mu}$ with  $n=(2,4,6,8,10)$ versus the Borel parameter $M^2$, where all input parameters are set to be their central values. In which, the shaded band indicate the Borel Windows for $n=(2,4,6,8,10)$, respectively.}
\label{fxinM2}
\end{figure}

\begin{table*}
\caption{Our predictions for the first five nonvanishing moments and inverse moment of the pion DA, compared to other theoretical predictions. Meanwhile, the values obtained by the formula combining Eq.~(\ref{xinxi0}) and Eq.~(\ref{moment0}) is also shown.}\label{table:xin_value}
\begin{tabular}{lccccccc}
\hline\hline
& ~~~$\mu{\rm [GeV]}$~~~ & ~~~$\langle\xi^2\rangle_{2;\pi}|_\mu$~~~ & ~~~$\langle\xi^4\rangle_{2;\pi}|_\mu$~~~ & ~~~$\langle\xi^6\rangle_{2;\pi}|_\mu$~~~ & ~~~$\langle\xi^8\rangle_{2;\pi}|_\mu$~~~ & ~~~$\langle\xi^{10}\rangle_{2;\pi}|_\mu$~~~&~~~$\langle x^{-1}\rangle|_\mu$ \\ \hline
BFTSR (This Work)               & 1  & 0.271(13) & 0.138(10)& 0.087(6)& 0.064(7)& 0.050(6)& 3.95\\
BFTSR (This Work)               & 2  & 0.254(10) & 0.125(7) & 0.077(6)& 0.054(5)& 0.041(4)& 3.33\\
Asymptotic              &$\infty$& 0.200 & 0.086    & 0.048   & 0.030   & 0.021   & 3.00\\
LF Holographic ($B=0$)~\cite{Ahmady:2018muv}  &1,2 & 0.180, 0.185 & 0.067, 0.071                             &-&-&-& 2.81,2.85\\
LF Holographic ($B\gg1$)~\cite{Ahmady:2018muv} &1,2 & 0.200, 0.200 & 0.085, 0.085                             &-&-&-& 2.93,2.95\\
LF Holographic~\cite{Brodsky:2007hb}  &$\sim 1$ & 0.237 & 0.114                                      &-&-&-& 4.0\\
Playkurtic~\cite{Stefanis:2014nla}         & 2  & $0.220^{+0.009}_{-0.006}$ & $0.098^{+0.008}_{-0.005}$   &-&-&-& $3.13^{+0.14}_{-0.10}$\\
LF Quark Model~\cite{Choi:2007yu}     & $\sim 1$ & 0.24(22) & 0.11(9)                                &-&-&-&-\\
Sum Rules~\cite{Ball:2004ye}          & 1  & 0.24 & 0.11                                             &-&-&-&-\\
Renormalon model~\cite{Agaev:2005rc}   & 1  & 0.28 & 0.13                                             &-&-&-&-\\
Instanton vacuum~\cite{Petrov:1998kg,Nam:2006au}& 1  & 0.22, 0.21  & 0.10,0.09                                 &-&-&-&-\\
NLC Sum Rules~\cite{Bakulev:2001pa}      & 2  & $0.248^{+0.016}_{-0.015}$ & $0.108^{+0.05}_{-0.03}$     &-&-&-& 3.16(9)\\
Sum Rules~\cite{Chernyak:1983ej}          & 2  & 0.343  & 0.181                                          &-&-&-& 4.25  \\
Dyson-Schwinger [RL,DB]~\cite{Chang:2013pq}  & 2 & 0.280, 0.251& 0.151, 0.128                          &-&-&-& 5.5,4.6 \\
Lattice~\cite{Arthur:2010xf}            & 2  & 0.28(1)(2) & -                                          &-&-&-&-\\
Lattice\cite{Braun:2015axa}           & 2  & 0.2361(41)(39)                & -                       &-&-&-&-\\
Lattice~\cite{Braun:2006dg}          & 2  & 0.27(4)                       & -                       &-&-&-&-\\
Lattice~\cite{Bali:2017ude}          & 2  & 0.2077(43)                       & -                       &-&-&-&-\\
Lattice~\cite{Bali:2019dqc}          & 2  & 0.234(6)(6)                       & -                       &-&-&-&-\\
Lattice~\cite{Zhang:2020gaj}          & 2  & 0.244(30)                       & -                       &-&-&-&-\\ \hline
Eq.~(\ref{xinxi0})+Eq.~(\ref{moment0})  & 1  & 0.303(19) & 0.179(21) & 0.128(16) & 0.098(14) & 0.082(20) &-\\ \hline\hline
\end{tabular}
\end{table*}

To study the influence of the Borel parameters to the pionic DA moments in the Borel window, we listed the results $\langle\xi^n\rangle_{2;\pi}|_{\mu}$ changed with Borel windows in Table~\ref{tbw}. In which the $\langle\xi^n\rangle_{2;\pi}|_{\mu}$ changed less than $10\%$ with the Borel windows, i.e. $4.1\%,~ 6.1\%,~ 8.6\%,~ 8.8\%,~7.1\%$ for $n=(2,4,6,8,10)$ respectively. Thus, the Borel windows for $\langle\xi^n\rangle_{2;\pi}|_{\mu}$ are stable to the BFTSR. Furthermore, the five curves of pionic leading-twist DA moments, i.e. $\langle\xi^n\rangle_{2;\pi}|_{\mu}$ for $n=(2,4,6,8,10)$ versus the Borel parameter $M^2$ are shown in Fig.~\ref{fxinM2}. The figure indicate that:
\begin{itemize}
  \item The curves for $\langle\xi^n\rangle_{2;\pi}|_{\mu}$ changed sharply in the small Borel area especially for the $M^2 \rightsquigarrow 1~{\rm GeV^2}$.
  \item The values of $\langle\xi^n\rangle_{2;\pi}|_{\mu}$ became small with the increasement for order $n$.
  \item The stable Borel parameter $M^2$ for $\langle\xi^n\rangle_{2;\pi}|_{\mu}$ become larger with the increase of $n$.
\end{itemize}

After taking all uncertainty sources into consideration, and adopting the RGE of moments mentioned in the above subsection, the first five nonvanishing values of $\langle\xi^n\rangle_{2;\pi}|_\mu$, i.e. $n=(2,4,6,8,10)$ within uncertainties coming from every input parameters are shown in Table~\ref{table:xin_value}. In which, the factorization scale are taken both the initial scale $\mu_0$ and typical scale $\mu = 2~{\rm GeV}$.  As a deeper comparison, we also listed the Light-front Holographic with $B=0$ and $B\gg 1$~\cite{Brodsky:2007hb,Ahmady:2018muv}, Playkurtic~\cite{Stefanis:2014nla}, LF Quark Model~\cite{Choi:2007yu}, QCD Sum Rules~\cite{Ball:2004ye, Chernyak:1983ej}, Renormalon model~\cite{Agaev:2005rc}, Instanton vacuum~\cite{Petrov:1998kg,Nam:2006au}, Non-Local Condensate (NLC) Sum Rules~\cite{Bakulev:2001pa}, Dyson-Schwinger [RL,DB]~\cite{Chang:2013pq}, Lattice~\cite{Arthur:2010xf, Braun:2015axa, Braun:2006dg,Bali:2017ude,Bali:2019dqc,Zhang:2020gaj}. At the same time, we also provide the inverse moment $\langle x^{-1}\rangle|_\mu = \int_0^1 dx x^{-1} \phi_{2;\pi}(x,\mu)$ in Table~\ref{table:xin_value}. In addition, in order to show the advantages of new sum rules formula (\ref{xin}), the values of $\langle\xi^n\rangle_{2;\pi}|_\mu$ obtained by the formula combining Eq.~(\ref{xinxi0}) and Eq.~(\ref{moment0}) commonly used in literature is also listed in this table. From the table, we can get the conclusions,

\begin{table*}[t]
\caption{Comparison of the second and fourth Gegenbauer moments of the pion leading-twist DA with different methods.}\label{table:GegenbauerMoment}
\begin{tabular}{lcccc}
\hline
 Method ~~~~~~~~~~~~~~~~~~~~ &~~~~~~~~~~$\mu({\rm GeV})$~~~~~~~~~~ & ~~~~~~~~~~~~~$a_2^{2;\pi}$~~~~~~~~~~~~~ &~~~~~~~~~~~~~$a_4^{2;\pi}$~~~~~~~~~~~~~\\ \hline
 BFTSR (This work)                               &1& $0.206\pm 0.038$ & $0.047\pm 0.011$  \\
 BFTSR (This work)                               &2& $0.157\pm 0.029$ & $0.032\pm 0.007$   \\
 Lattice~\cite{Arthur:2010xf}                       &2& $0.233\pm 0.065$ &   \\
 Lattice~\cite{Braun:2015axa}                       &2& $0.136\pm 0.021$ &   \\
 Lattice~\cite{Bali:2019dqc}                        &1& $0.135\pm 0.032$ &   \\
 Lattice~\cite{Bali:2019dqc}                        &2& $0.101\pm 0.023$ &   \\
 Sum rules~\cite{Mikhailov:2016klg,Stefanis:2020rnd} &1& $0.203^{+0.069}_{-0.057}$ & $-0.143^{+0.094}_{-0.087}$  \\
 Sum rules~\cite{Mikhailov:2016klg,Stefanis:2020rnd} &2& $0.149^{+0.052}_{-0.043}$ & $-0.096^{+0.063}_{-0.058}$  \\
 LCSR fitting~\cite{Khodjamirian:2011ub}         &1& $0.17\pm 0.08$ & $0.06\pm 0.10$  \\
 LCSR fitting~\cite{Agaev:2010aq}                &2& $0.096$ &   \\
 LCSR fitting~\cite{Agaev:2012tm}                &2& $0.067$ &   \\
 LCSR fitting~\cite{Bruschini:2020voj}           &1& $0.22-0.33$ & $0.12-0.25$  \\
 Dyson-Schwinger (RL)~\cite{Chang:2013pq}                     &2& $$0.233 & $0.115$  \\
 Dyson-Schwinger (DB)~\cite{Chang:2013pq}                     &2& $0.149$ & $0.080$  \\ \hline
\end{tabular}
\end{table*}
\begin{figure*}[t]
\centering
\includegraphics[width=0.33\textwidth]{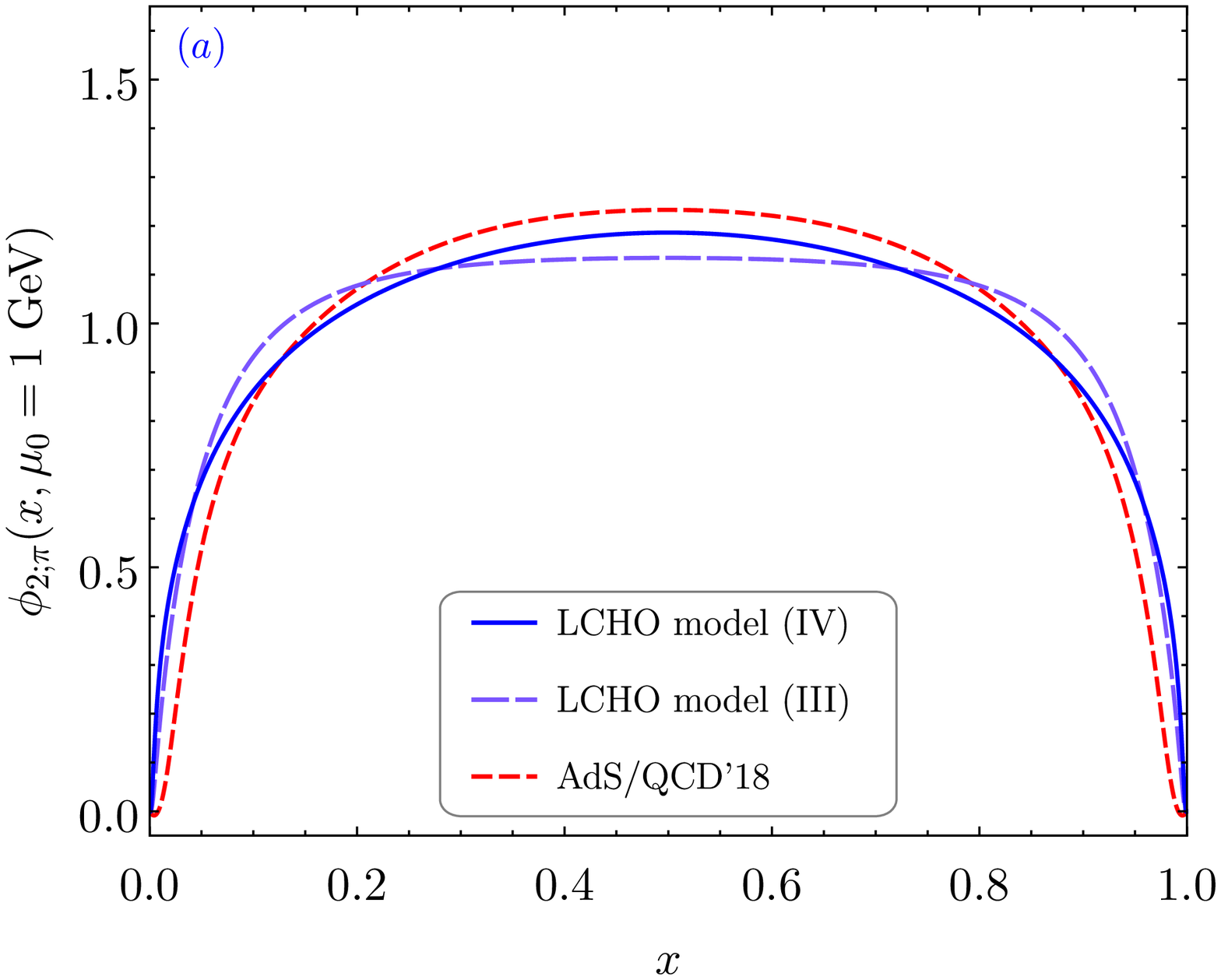}\includegraphics[width=0.33\textwidth]{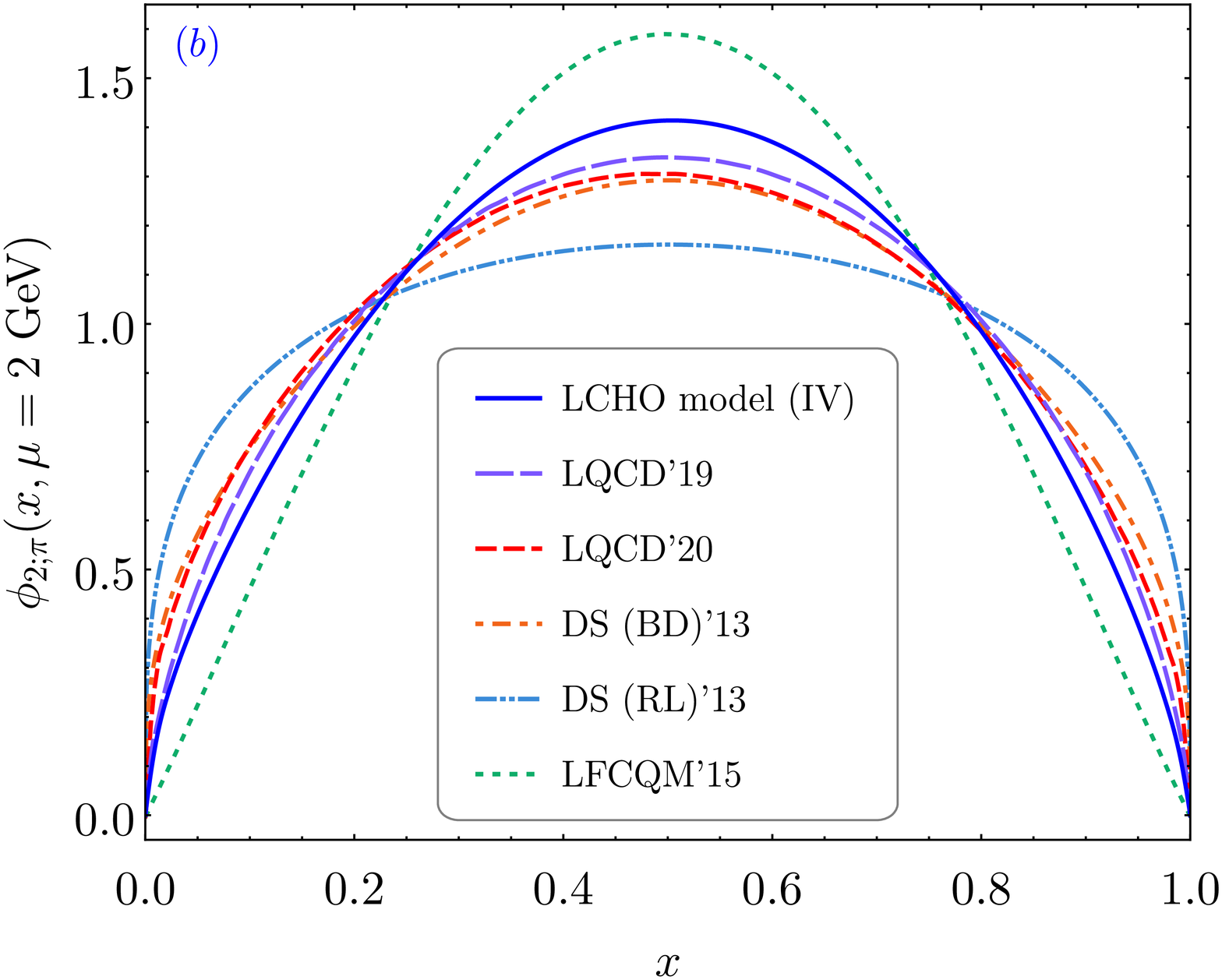}\includegraphics[width=0.33\textwidth]{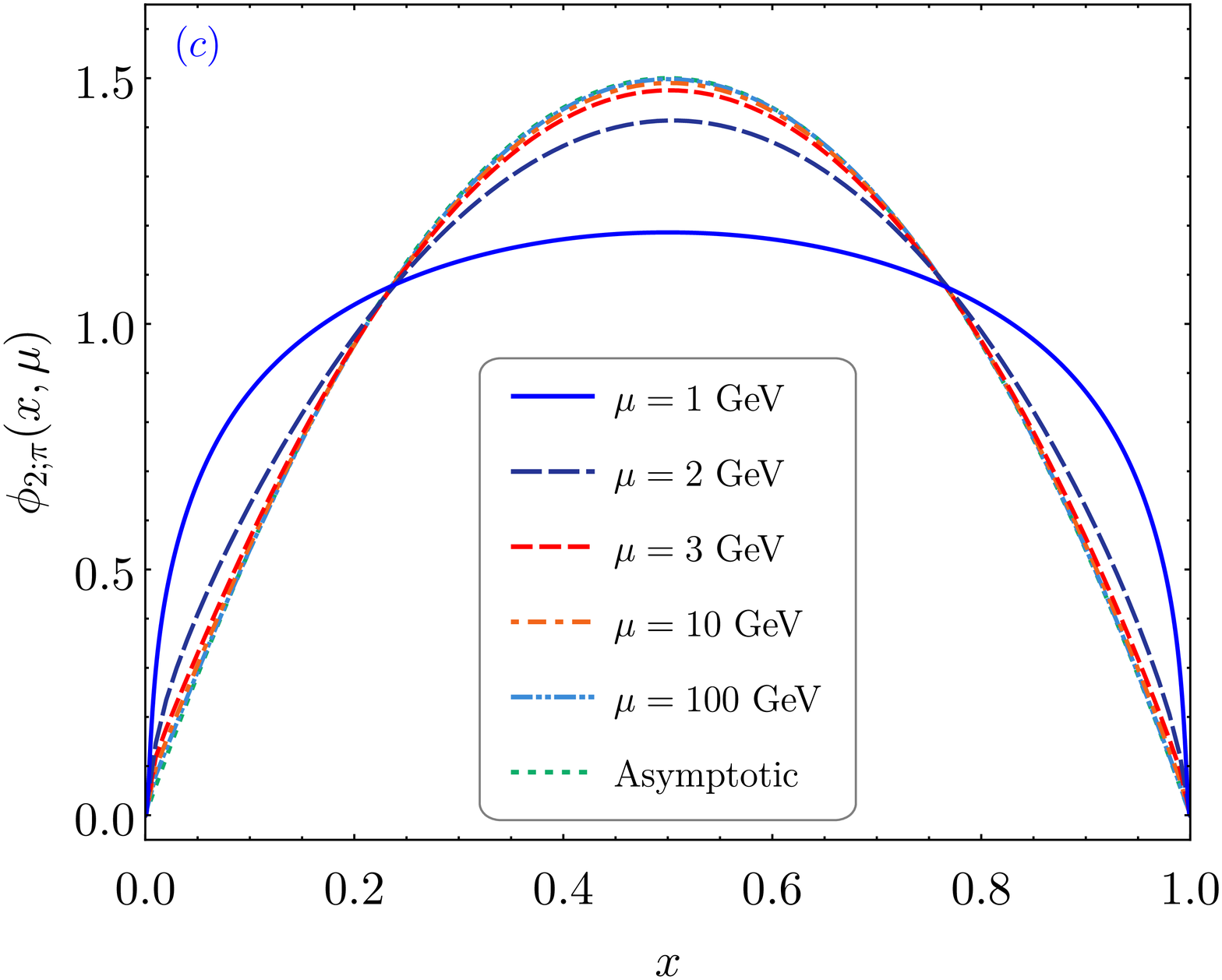}
\caption{The pionic leading-twist DA curves in this work. For the panel-(a) and panel-(b), we present DS model~\cite{Chang:2013pq}, QCD/AdS model~\cite{Ahmady:2018muv}, the DAs by LFCQM\cite{deMelo:2015yxk} and LQCD~\cite{Bali:2019dqc,Zhang:2020gaj} as a comparison. For the panel-(c), our LCHO model-IV at several typical energy scales e.g., $\mu =1, 2, 3, 10, 100~{\rm GeV}$ are given respectively.}
\label{fDAcures}
\end{figure*}
\begin{itemize}
  \item Up to 10th-order accuracy, we provide a complete series results for $\langle\xi^n\rangle_{2;\pi}|_\mu$ within uncertainties.
  \item For the $n=(2,4)$ cases, our results have a good agreement with the DS model and Lattice results.
  \item The inverse moment at $\mu=2~{\rm GeV}$ of our prediction is closely to the  Playkurtic and NLC Sum Rules  results.
  \item Comparing the values in the first and last row, one can find that the differences between corresponding moments are about $12\%$, $30\%$, $47\%$, $53\%$ and $64\%$ for $n = 2, 4, 6, 8, 10$, respectively. These ratios can be regarded as the accuracy improved by adopting new sum rules formula (\ref{xin}). At the same time, one can find that those differences increased with the increase of the order-$n$. The reason is that the Borel window moves to the right with the increase of order-$n$ (see Table \ref{tbw}), and the deviation of the sum rule of $0^{\rm th}$ moment, Eq.~(\ref{xi0xi0}), from normalization, increases with the increase of Borel parameter. The errors in the first row are significantly less than that in the last row. The reason is that the sum rules (\ref{xin}) can eliminate some systematic errors caused by the selection and determination of various input parameters. To calculate $\langle\xi^n\rangle_{2;\pi}|_\mu$ by combining Eq.~(\ref{xinxi0}) and Eq.~(\ref{moment0}), we have required that the continuum state contributions are less than $45\%$, $50\%$, $50\%$, $55\%$, $55\%$; and the dimension-six contributions are not more than $10\%$, $15\%$, $15\%$, $15\%$, $15\%$, for the order-$n = 2, 4, 6, 8, 10$, respectively. Comparing the criterions adopted for sum rules (\ref{xin}) mentioned above, which are obviously much larger. This means that the sum rules (\ref{xin}) does eliminate some systematic errors caused by the continuum state and the absence of high dimensional condensates.
\end{itemize}
Moreover, considering the low reliability of high order Gegenbauer moments, we only give the values of the second and forth Gegenbauer moments in this paper, which are shown in Table~\ref{table:GegenbauerMoment}.As a comparision, the values by QCD sum rules~\cite{Mikhailov:2016klg,Stefanis:2020rnd}, Lattice~\cite{Arthur:2010xf,Braun:2015axa,Bali:2019dqc}, LCSR fitting~\cite{Khodjamirian:2011ub,Agaev:2010aq,Agaev:2012tm,Bruschini:2020voj} and Dyson-Schwinger [RL,DB]~\cite{Chang:2013pq} are also present. In which, our predictions have agreement with the QCD sum rules, LCSR fitting and the Dyson-Schwinger equations predictions within errors.

\subsection{The model parameters of the pionic leading-twist DA and applications}

Combining the the normalization condition (\ref{DA_constraint1}) and the sum rule (\ref{DA_constraint2}) derived from $\pi^0\to\gamma\gamma$ decay amplitude, making use of the least square method mentioned in Sec.II to fit the values of moments $\langle\xi^n\rangle_{2;\pi}|_{\mu}$ shown in Table~\ref{table:xin_value}, the parameters of our LCHO model-III can be obtained:
\begin{eqnarray}
A_{2;\pi}         &=& 14.7999 {\rm GeV}^{-1}, \nonumber\\
\alpha_{2;\pi}    &=& -0.158,                 \nonumber\\
\beta_{2;\pi}     &=& 0.920029 {\rm GeV},
\label{ModelParameterIII}
\end{eqnarray}
with $\chi^2_{\rm min}/n_d = 0.437236/4$, $P_{\chi^2_{\rm min}} = 0.979316$. The parameters of our LCHO model-IV are:
\begin{eqnarray}
A_{2;\pi}         &=& 5.95481 {\rm GeV}^{-1}, \nonumber\\
\alpha_{2;\pi}    &=& -0.717,                 \nonumber\\
\hat{a}^{2;\pi}_2 &=& -0.125,                   \nonumber\\
\beta_{2;\pi}     &=& 0.937482 {\rm GeV},
\label{ModelParameterIV}
\end{eqnarray}
with $\chi^2_{\rm min}/n_d = 0.119251/3$, $P_{\chi^2_{\rm min}} = 0.989431$.

The curves of our prediction is shown in Figure~\ref{fDAcures}. For comparison, DS model~\cite{Chang:2013pq}, QCD/AdS model with $B=1$~\cite{Ahmady:2018muv}, the DAs by the light-front constituent quark model (LFCQM)\cite{deMelo:2015yxk} and LQCD~\cite{Bali:2019dqc,Zhang:2020gaj} are also shown in Figure~\ref{fDAcures}.

\begin{itemize}
\item  From the panel-(a) in Figure~\ref{fDAcures}, one can find that our LCHO model-III is near flat in region $x\in [0.2,0.8]$, and is a little wider than LCHO model-IV, both of them very close to the AdS/QCD model. With the model parameters of LCHO model-IV in Eq.~(\ref{ModelParameterIV}), one can calculate the moments of DA, i.e. $\langle \xi^n \rangle_{2;\pi}^{\rm IV} |_{\mu_0} = (0.269, 0.140, 0.089, 0.063, 0.048)$ for $n=(2,4,6,8,10)$, respectively. Those values are also very closely to the references results in Table~\ref{table:xin_value}.
\item By substituting our LCHO model-III with parameters in Eqs.~(\ref{ModelParameterIII}) into Eq.~(\ref{moment}), we can get $\langle \xi^n \rangle_{2;\pi}^{\rm III} |_{\mu_0} = (0.275, 0.142, 0.089, 0.062, 0.046)$ for $n=(2,4,6,8,10)$, respectively. Compared our LCHO model-III with LCHO model-IV, the later is better, which will be used in the following discussion and calculation, and omitted the mark ``IV''.
\item  From the panel-(b) in Figure~\ref{fDAcures}, one can find that, our LCHO model is narrower than DS model, wider than that by LFCQM, and closer to the LQCD result in Ref.~\cite{Bali:2019dqc}.
\end{itemize}

The pionic twist-2 DA behavior of our model at any other scale can be related to that of in initial scale by using the energy evolution equation~\cite{Huang:2013yya}, which are shown in the panel-(c) of Figure~\ref{fDAcures}. One can find that,
\begin{itemize}
  \item Our LCHO model at $\mu_0$ is significantly broader than the asymptotic form.
  \item With the increase of scale $\mu$, our pionic leading-twist DA model curve becomes narrower and closer to the asymptotic form. Especially, when the scale $\mu$ is lower than $2~{\rm GeV}$, our pionic leading-twist DA behavior is more sensitive to $\mu$, while when $\mu > 2~{\rm GeV}$, which is close to the asymptotic behavior and insensitive to the scale $\mu$.
  \item
      In order to have a clear look at the changes of LCDA with factorization scale, one can set $x = 0.5$ and numerical results are $\phi_{2;\pi}^{\rm IV}(x=0.5,\mu) = (1.186, 1.414, 1.475, 1.490, 1.498)$ for $\mu = (1, 2, 3, 10, 100)~{\rm GeV}$ respectively.
\end{itemize}

\begin{table*}[t]
\caption{The model parameters of our LCHO model with $\varphi_{2;\pi}^{\rm IV}(x)$ and the corresponding goodness-of-fit for several typical constituent quark mass $m_q$.}\label{table:fitting}
\begin{tabular}{c c c c c c}
\hline
~~~$m_q({\rm MeV})$~~~ & ~~~$A_{2;\pi}(\rm GeV^{-1})$~~~ & ~~~$\alpha_{2;\pi}$~~~ & ~~~$B_2^{2;\pi}$~~~ & ~~~$\beta_{2;\pi}({\rm GeV})$~~~ & ~~~$P_{\chi^2_{\rm min}}$ \\ \hline
~~~$350$~~~ & ~~~$2.24732$~~~ & ~~~$-1.382$~~~ & ~~~$-0.115$~~~ & ~~~$0.608317$~~~ & ~~~$0.797900$ \\
~~~$340$~~~ & ~~~$2.40934$~~~ & ~~~$-1.330$~~~ & ~~~$-0.115$~~~ & ~~~$0.617249$~~~ & ~~~$0.824339$ \\
~~~$330$~~~ & ~~~$2.55114$~~~ & ~~~$-1.286$~~~ & ~~~$-0.116$~~~ & ~~~$0.627146$~~~ & ~~~$0.848822$ \\
~~~$320$~~~ & ~~~$2.70534$~~~ & ~~~$-1.242$~~~ & ~~~$-0.117$~~~ & ~~~$0.637809$~~~ & ~~~$0.871190$ \\
~~~$310$~~~ & ~~~$2.92676$~~~ & ~~~$-1.186$~~~ & ~~~$-0.116$~~~ & ~~~$0.649553$~~~ & ~~~$0.891410$ \\
~~~$300$~~~ & ~~~$3.10553$~~~ & ~~~$-1.143$~~~ & ~~~$-0.117$~~~ & ~~~$0.662378$~~~ & ~~~$0.909415$ \\
~~~$290$~~~ & ~~~$3.32326$~~~ & ~~~$-1.095$~~~ & ~~~$-0.117$~~~ & ~~~$0.676766$~~~ & ~~~$0.925250$ \\
~~~$280$~~~ & ~~~$3.52839$~~~ & ~~~$-1.053$~~~ & ~~~$-0.118$~~~ & ~~~$0.692543$~~~ & ~~~$0.938962$ \\
~~~$270$~~~ & ~~~$3.82177$~~~ & ~~~$-0.999$~~~ & ~~~$-0.117$~~~ & ~~~$0.710313$~~~ & ~~~$0.950620$ \\
~~~$260$~~~ & ~~~$4.14554$~~~ & ~~~$-0.945$~~~ & ~~~$-0.116$~~~ & ~~~$0.730185$~~~ & ~~~$0.960338$ \\
~~~$250$~~~ & ~~~$4.45116$~~~ & ~~~$-0.898$~~~ & ~~~$-0.116$~~~ & ~~~$0.752879$~~~ & ~~~$0.968389$ \\
~~~$240$~~~ & ~~~$4.79603$~~~ & ~~~$-0.850$~~~ & ~~~$-0.116$~~~ & ~~~$0.778565$~~~ & ~~~$0.974850$ \\
~~~$230$~~~ & ~~~$5.21220$~~~ & ~~~$-0.797$~~~ & ~~~$-0.115$~~~ & ~~~$0.808690$~~~ & ~~~$0.979973$ \\
~~~$220$~~~ & ~~~$5.63027$~~~ & ~~~$-0.749$~~~ & ~~~$-0.115$~~~ & ~~~$0.843751$~~~ & ~~~$0.983961$ \\
~~~$210$~~~ & ~~~$5.85107$~~~ & ~~~$-0.726$~~~ & ~~~$-0.119$~~~ & ~~~$0.885935$~~~ & ~~~$0.987168$ \\
~~~$200$~~~ & ~~~$5.95481$~~~ & ~~~$-0.717$~~~ & ~~~$-0.125$~~~ & ~~~$0.937482$~~~ & ~~~$0.989431$ \\ \hline
\end{tabular}
\end{table*}

\begin{figure}[h]
\centering
\includegraphics[width=0.45\textwidth]{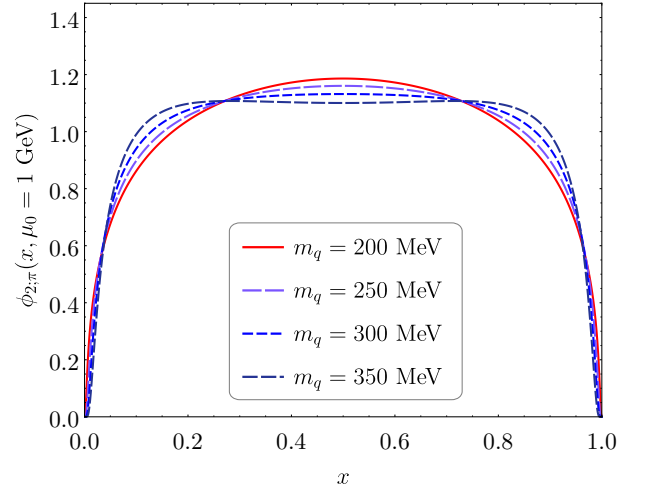}
\caption{The curves of our LCHO model for the pionic leading-twist DA $\phi_{2;\pi}(x,\mu)$ at $\mu = 1{\rm GeV}$ with the constituent quark mass $m_q = 200, 250, 300, 350~{\rm MeV}$, respectively.}
\label{DAcuresmq}
\end{figure}

As a further step, the sensitivity/goodness-of-fit for the behavior of our LCHO model $\varphi_{2;\pi}^{\rm IV}(x)$ with the constituent quark mass, i.e. $m_q = (350,340,\cdots,200){\rm GeV}$ are also been analyzed exhibited in Table~\ref{table:fitting}, which indicate the value of goodness-of-fit increasing with the decrease of constituent quark mass. The $P_{\chi^2_{\rm min}}$ will less than 0.9 when $m_q > 300~{\rm MeV}$. In order to more intuitively understand the impact of $m_q$ on our $\phi_{2;\pi}(x,\mu)$, the curves of our LCHO model for the pionic leading-twist DA $\phi_{2;\pi}(x,\mu)$ at $\mu = 1{\rm GeV}$ with the constituent quark mass $m_q = (200, 250, 300, 350)~{\rm MeV}$ are shown in Figure~\ref{DAcuresmq}. One can find that, with the increase of $m_q$, our model tends to the flat-like form.

Within the resultant LCHO model of our predictions, there also exist Jacobi factor $\sqrt{{\partial k_n}/{\partial x}}$ contribute to the wave functions~\cite{Choi:1997iq}, which can be read off,
\begin{eqnarray}
\frac{\partial k_n}{\partial x} = \frac{M_0}{4x\bar x} \bigg[ 1 - \bigg(\frac{m_q^2 - m_{\bar{q}}^2}{M_0^2} \bigg)^2 \bigg],
\end{eqnarray}
with $M_0^2 = (\textbf{k}_\perp^2 + m_q^2)/x + (\textbf{k}^2_\perp + m_{\bar{q}}^2)/\bar x$. Due to the invariant meson mass scheme~\cite{Terentev:1976jk, Jaus:1989au, Jaus:1991cy, Chung:1988mu,Choi:1997qh,Schlumpf:1994bc, Cardarelli:1994yq}, and one can take $m_q = m_{\bar q}$ for pion cases, then the spatial wave function would be
\begin{eqnarray}
\Psi_{2;\pi}^R(x,\textbf{k}_\perp) = A_{2;\pi} \varphi_{2;\pi}(x) \frac{\sqrt{\textbf{k}_\perp^2 + m_q^2}}{4 (x\bar x)^{3/2}} \exp \left[ - \frac{\textbf{k}_\perp^2 + m_q^2}{8\beta_{2;\pi}^2 x\bar x} \right],\nonumber\\
\end{eqnarray}
Finally, we can get the expression of pionic twist-2 LCDA
\begin{align}
\phi_{2;\pi}(x,\mu) &= \frac{\sqrt{3/2} A_{2;\pi} m_q \beta_{2;\pi}^2 \varphi_{2;\pi}(x)}{2\pi^2 f_\pi \sqrt{x\bar x}} \exp \left[ - \frac{m_q^2}{8\beta_{2;\pi}^2 x\bar x} \right] \nonumber\\
&\times \left\{ 1 - \exp \left[ - \frac{\mu^2}{8\beta_{2;\pi}^2 x\bar x} \right] \right\}
\end{align}
\begin{figure*}[htb]
\centering
\includegraphics[width=0.33\textwidth]{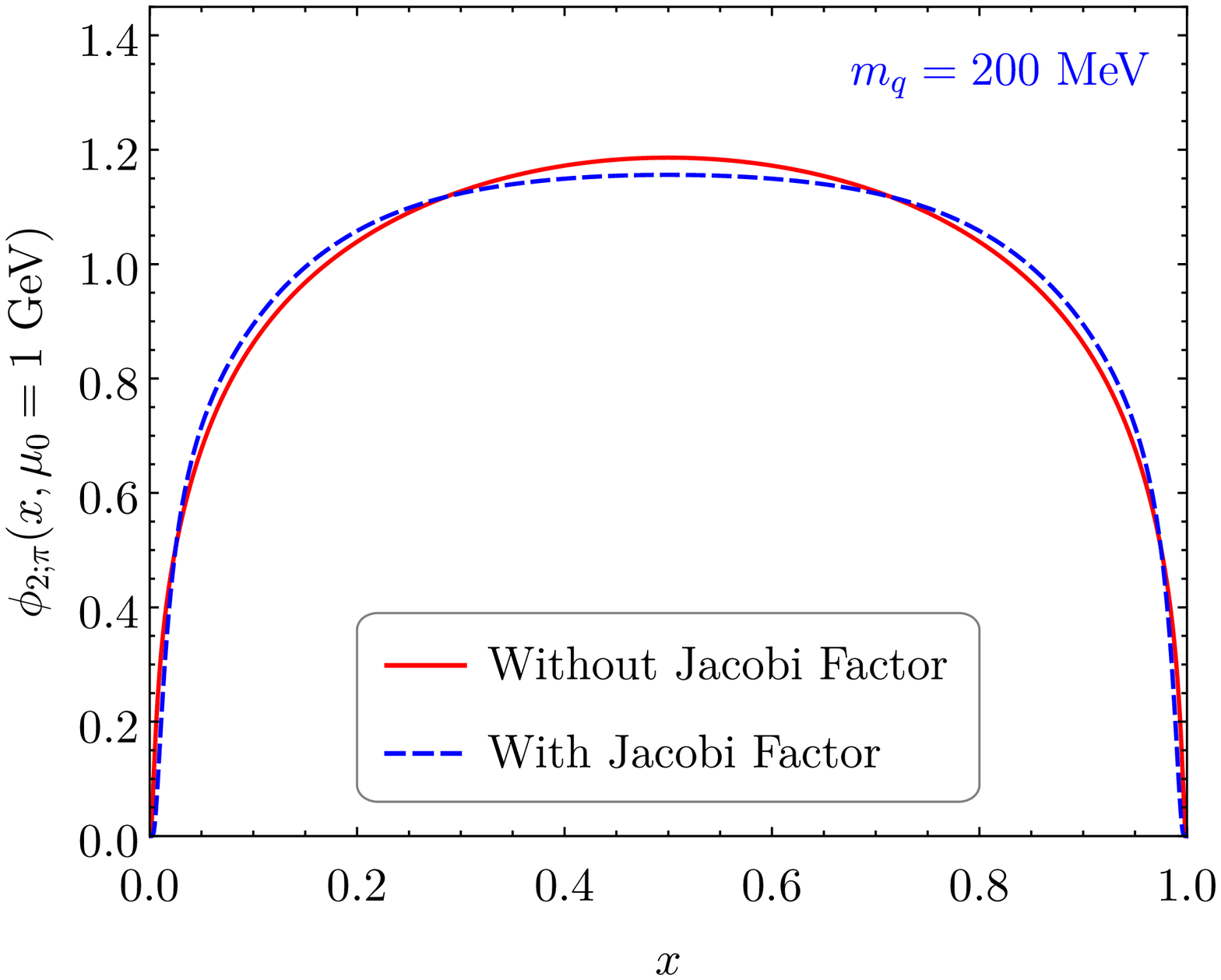}\includegraphics[width=0.33\textwidth]{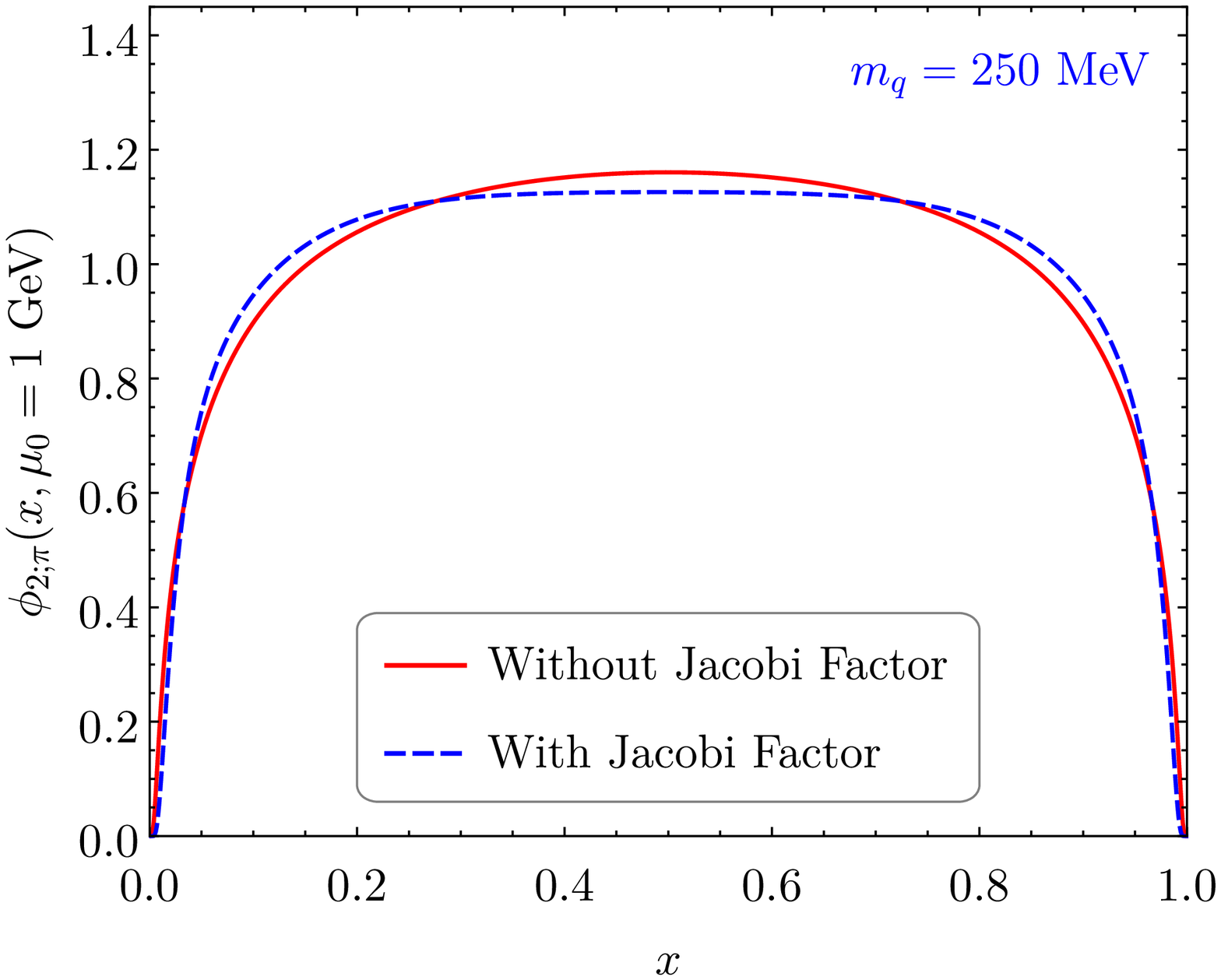}\includegraphics[width=0.33\textwidth]{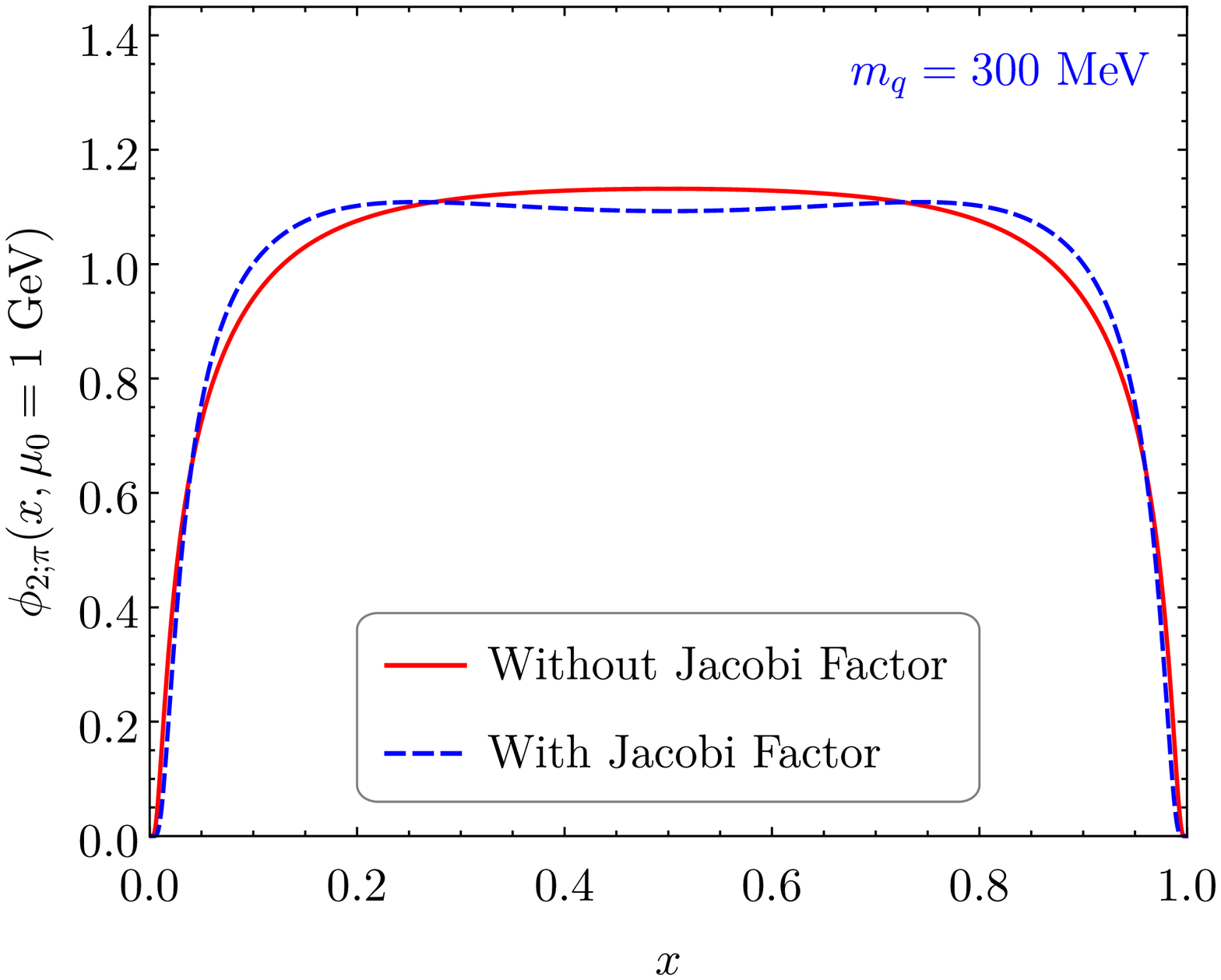}
\caption{Comparison of two pionic leading-twist DA LCHO models with and without the Jacobi factor. Three plans are correspond to the constituent quark mass $m_q = 200, 250, 300~{\rm MeV}$ respectively.}
\label{DAcuresjacobimq}
\end{figure*}
Then we can fit the values of the moments $\langle\xi^n\rangle_{2;\pi}$ from the sum rules Eq.~\eqref{xin}, by using the least square method with the above model. Comparing the behavior of the two pionic leading-twist DA LCHO model with the Jacobi factor and without the Jacobi factor, the difference between the two is not obvious, which are also shown in Figure \ref{DAcuresjacobimq}.\\

\begin{figure}[t]
\centering
\includegraphics[width=0.42\textwidth]{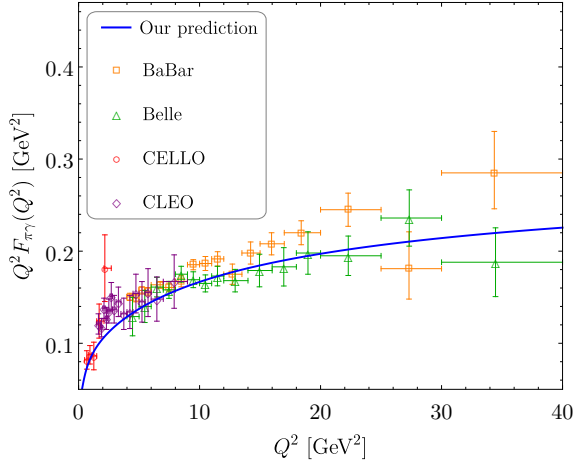}
\caption{The pion-photon TFF $Q^2F^{(V)}_{\pi\gamma}(Q^2)$ with our model. For comparison, the experimental data reported by CELLO~\cite{Behrend:1990sr}, CLEO~\cite{CLEO,Gronberg:1997fj}, BaBar~\cite{Aubert:2009mc} and Belle~\cite{Uehara:2012ag} collaborations are shown.}
\label{fPionPhotonTFF}
\end{figure}

As significant applications, we recalculate the pion-photon TFF $F_{\pi\gamma}(Q^2)$ and the $B\to\pi$ TFF $f^{B\to\pi}_+(q^2)$ with our pionic leading-twist DA model. The pion-photon TFF $F_{\pi\gamma}(Q^2)$ can be calculated with LCSR~\cite{Mikhailov:2016klg,Stefanis:2020rnd,Mikhailov:2021znq} and pQCD method~\cite{Wu:2010zc,Huang:2006wt}. With pQCD method, $F_{\pi\gamma}(Q^2)$ can be expressed as the sum of the valence quark part contribution $F^{(\rm V)}_{\pi\gamma}(Q^2)$ and the non-valence quark part contribution $F^{({\rm NV})}_{\pi\gamma}(Q^2)$,
\begin{eqnarray}
F_{\pi\gamma}(Q^2) = F^{(\rm V)}_{\pi\gamma}(Q^2) + F^{(\rm NV)}_{\pi\gamma}(Q^2),
\label{PionPhotonTFF}
\end{eqnarray}
where the corresponding analytical formula of $F^{(\rm V)}_{\pi\gamma}(Q^2)$ and $F^{({\rm NV})}_{\pi\gamma}(Q^2)$ can be found in Refs.~\cite{Wu:2010zc,Huang:2006wt}. Figure~\ref{fPionPhotonTFF} show the curve of $Q^2 F_{\pi\gamma}(Q^2)$ versus $Q^2$ by our pionic leading-twist DA model and the experimental data reported by CELLO\cite{Behrend:1990sr}, CLEO~\cite{CLEO,Gronberg:1997fj}, BaBar~\cite{Aubert:2009mc} and Belle~\cite{Uehara:2012ag} collaborations, and one can find that our prediction is consistent with the BELLE data in large $Q^2$ region.

Furthermore, as another important application for the pion twist-2 LCDA, the TFF for the $B\to\pi$ decay processes should be analysis. We start with the following correlation function
\begin{align}
\Pi_\mu(p,q) = i\int d^4x e^{iq\cdot x}\langle\pi^+(p)|T \{j_V^\mu(x),j_B^\dag(0)\} |0\rangle
\end{align}
with $j_V^\mu(x) = \bar u(x)\gamma_\mu(1+\gamma_5)b(x)$. For the current of $B$-meson $j_B^\dag(0)$, we choice the right-handed current $j_B^\dag(0)=m_b b(0) i(1+\gamma_5) d(0)$ which can highlight the twist-2, 4 DAs contributions, and the twist-3 DAs contributions vanished. By following the standard procedures of light-cone sum rule approach~\cite{Huang:2001xb,Duplancic:2008ix}, we can get the $B\to\pi$ TFF $f_+^{B\to\pi}(q^2)$, reads
\begin{eqnarray}
f_+^{B\to\pi}(q^2) &=& \frac{e^{m_B^2/M^2}}{m_B^2 f_B} \left[ F_0(q^2, M^2, s_0^B) \right. \nonumber\\
&+& \left. \frac{\alpha_s C_F}{4\pi} F_1(q^2, M^2, s_0^B) \right],
\label{BtoPionTFF}
\end{eqnarray}
where $C_F = 4/3$, $m_B$ and $f_B$ are the $B$-meson mass and decay constant respectively, $s_0^B$ is the continuum threshold. The LO contribution of the LCSR (\ref{BtoPionTFF}) is expressed as
\begin{eqnarray}
&&F_0(q^2, M^2, s_0^B)
\nonumber\\
&& \qquad = m_b^2 f_\pi \int^1_{u_0} du e^{-\frac{m_b^2-q^2\bar{u}}{uM^2}} \left\{ \frac{\phi_{2;\pi}(u)}{u} + \frac{1}{m_b^2-q^2} \right.
\nonumber\\
&& \qquad \times \left[ -\frac{m_b^2u}{4(m_b^2-q^2)} \frac{d^2\phi_{4\pi}(u)}{du^2} + u\psi_{4\pi}(u) \right.
\nonumber\\
&& \qquad + \left.\left. \int^u_0 dv \psi_{4\pi}(v) - I_{4\pi}(u) \right] \right\},
\label{F0}
\end{eqnarray}
and the NLO term of $f^{B\to\pi}_+(q^2)$ is
\begin{align}
&F_1(q^2, M^2, s_0^B)
\nonumber\\
&\qquad = \frac{f_\pi}{\pi} \int^{s_0^B}_{m_b^2} ds e^{-s/M^2} \int^1_0 du {\rm Im}_s T_1(q^2,s,u) \phi_{2;\pi}(u).
\label{F1}
\end{align}
Where $m_b$ is the $b$-quark mass, $\bar{u} = 1-u$, $u_0 = (m_b^2-q^2)/(s_0^B-q^2)$, $\phi_{4\pi}(u)$ and $\psi_{4\pi}(u)$ are the pionic twist-4 DAs, and $I_{4\pi}(u)$ is the combination function of four pionic twist-4 DAs $\Psi_{4\pi}(u)$, $\Phi_{4\pi}(u)$, $\widetilde{\Psi}_{4\pi}(u)$ and $\widetilde{\Phi}_{4\pi}(u)$. For the expressions of those pionic twist-4 DAs, $I_{4\pi}(u)$, and the imaginary part of the amplitude $T_1$, one can find in Ref.~\cite{Duplancic:2008ix}. By taking $\mu = 3{\rm GeV}$, $M^2 = 18\pm 1{\rm GeV}^2$, $s_0^B = 35.75 \pm 0.25 {\rm GeV}^2$, $m_B = 5.279{\rm GeV}$, $f_B = 214_{-5}^{+7}{\rm MeV}$~\cite{Duplancic:2008ix}, we can obtain
\begin{align}
f_+^{B\to\pi}(0) = 0.295^{+0.018}_{-0.013}.
\label{fBpi}
\end{align}
This value is consistent with other theoretical group Refs.~\cite{Duplancic:2008ix,Li:2012gr, Imsong:2014oqa,Khodjamirian:2017fxg} by the conventional current correlation. The difference between the central value in Eq. (\ref{fBpi}) and the one in Ref.~\cite{Duplancic:2008ix} is mainly due to the difference in the selected correlation function. Comparing Eqs. \eqref{BtoPionTFF} - \eqref{F1} above with Eqs. (4.4), (4.5), (4.7) in Ref.~\cite{Duplancic:2008ix}, one can find that the contributions from pionic twist-3 DAs disappeared, while the contributions of pionic twist-2,4 DAs doubled. Then the difference between the twist-2 DA's contribution and twist-3 DAs' contributions in the LCSR with the conventional current correlation can be used as the system error caused by adopted the chiral current correlation function.

\section{summary}\label{Sec.V}
In this paper, we have improved the traditional LCHO model of pionic leading-twist DA $\phi_{2;\pi}(x,\mu)$ by introducing a new WF's longitudinal DA, i.e., $\varphi_{2;\pi}^{\rm IV}$ in Eq.~(\ref{varphiIV}). At the same time, we have improved the method of determining the model parameters. More explicitly, the least square method is adopted to fit the moments $\langle\xi^n\rangle_{2;\pi}|_{\mu}$ directly to determine the model parameters. This makes it necessary and meaningful to calculate higher-order moments. And we can obtain a stronger constraint on the DA behavior by including more moments.

We have adopted the QCD sum rules based on the BFT to calculate the moments $\langle\xi^n\rangle_{2;\pi}|_{\mu}$, and the values of first five moments are $\langle \xi^2\rangle_{2;\pi}|_{\mu_0}  = 0.271 \pm 0.013$, $\langle \xi^4\rangle_{2;\pi}|_{\mu_0}  = 0.138 \pm 0.010$, $\langle \xi^6\rangle_{2;\pi}|_{\mu_0}  = 0.087 \pm 0.008$, $\langle \xi^8\rangle_{2;\pi}|_{\mu_0}  = 0.064 \pm 0.007$, $\langle \xi^{10}\rangle_{2;\pi}|_{\mu_0}  = 0.050 \pm 0.006$, respectively. Based on those values, we obtain the behavior of $\phi_{2;\pi}(x,\mu)$, that is, Eqs.~\eqref{DA_model}, \eqref{varphiIV} and \eqref{ModelParameterIV}.

Compared with our previous work, in addition to the improvement of the LCHO model, there are three improvements: i) The moments $\langle\xi^n\rangle_{2;\pi}|_{\mu}$, rather than the Gegenbauer moments $a^{2;\pi}_n(\mu)$, are used as constraint conditions to determine the model parameters; ii) The least square method is used to fit the moments $\langle\xi^n\rangle_{2;\pi}|_{\mu}$ to get the appropriate model parameters; iii) We take Eq.~(\ref{xin}) rather than Eq.~(\ref{xinxi0}) as the sum rules of $\langle\xi^n\rangle_{2;\pi}|_{\mu}$, which can avoid the error caused by non normalized $0^{\rm th}$ moment $\langle\xi^0\rangle_{2;\pi}|_{\mu}$ on the left side of Eq.~(\ref{xinxi0}), and make the accuracy of the resulted values of  $\langle \xi^n \rangle_{2;\pi} |_{\mu}$ to be increased by more than $10\%$. Those improvements can be widely used to QCD sum rules studies of other meson DA to obtain more accurate DA's behavior.

As an application, we have taken our model to calculate the pion-photon TFF $F_{\pi\gamma}(Q^2)$ which are shown in Figure~\ref{fPionPhotonTFF}. Our results agree with the Belle predictions at large $Q^2$-region. Meanwhile, the $B\to\pi$ TFF $f^{B\to\pi}_+(q^2)$ has been calculated up to NLO accuracy, which agrees with other theoretical predictions.

\section{Acknowledgments}
This work was supported in part by the National Natural Science Foundation of China under Grant No.11765007, No.11875122, No.11625520, No.11947406 and No.12047564, the Project of Guizhou Provincial Department of Science and Technology under Grant No.KY[2019]1171, and No.ZK[2021]024, the Project of Guizhou Provincial Department of Education under Grant No.KY[2021]030, the Chongqing Graduate Research and Innovation Foundation under Grant No.ydstd1912, the Fundamental Research Funds for the Central Universities under Grant No.2020CQJQY-Z003, the Project of Guizhou Minzu University under Grant No. GZMU[2019]YB19.

\end{document}